\documentclass[acmtog]{acmart}
\usepackage[utf8x]{inputenc}
\usepackage{graphicx}
\usepackage{tikz}
\usepackage{multicol}
\usepackage[listings,skins]{tcolorbox}
\usepackage[inline]{enumitem} 
\usepackage{flushend}
\usepackage{caption}

\lstdefinestyle{nccode}{
    numbers=none,
    firstnumber=1,
    stepnumber=1,
    numbersep=10pt,
    tabsize=4, 
    showspaces=false,
    breaklines=true, 
    showstringspaces=false,
    moredelim=**[is][\btHL]{`}{`},
    moredelim=**[is][{\btHL[fill=black!10]}]{`}{`},
    moredelim=**[is][{\btHL[fill=celadon!40]}]{!}{!}
}

\usepackage{xspace}
\usepackage{soul}
\usepackage[titles]{tocloft}
\usepackage{array}
\usepackage{longtable}
\usepackage{nameref}
\usepackage{varwidth}
\usepackage{hyperref}  
\usepackage{svg}

\newcommand{\repourl}{\url{https://github.com/Jacarte/MEWE}\xspace}

\makeatother

\usepackage{nameref}
\usepackage{varwidth} %for the varwidth minipage environment

\definecolor{commentgreen}{RGB}{176, 176, 176}
\definecolor{rowcolor}{cmyk}{0,0.87,0.68,0.32}
\definecolor{rowcolor2}{cmyk}{ 20, 0, 37, 34}

\definecolor{eminence}{RGB}{108,48,130}
\definecolor{weborange}{RGB}{255,165,0}
\definecolor{frenchplum}{RGB}{129,20,82}
\definecolor{darkgreen}{RGB}{10, 92, 10}

\newcommand*\step[1]{
\noindent\tikz[baseline=(char.base)]{
        \node[shape=circle,text=black,draw=black, fill=white,inner sep=1.2pt] (char) {#1};}}

\definecolor{celadon}{rgb}{0.67, 0.88, 0.69}
\newcolumntype{a}{>{\columncolor{rowcolor!40}}r}
\newcolumntype{t}{>{\columncolor{celadon!50}}r}
\newcolumntype{g}{>{\columncolor{gray!30}}r}

\usepackage{newfloat}
\DeclareFloatingEnvironment[fileext=frm,placement={tph},name=Listing]{code}
\lstdefinestyle{WATStyle}{
  numbers=left,
  stepnumber=1,
  numbersep=10pt,
  tabsize=4,
  showspaces=false,
  showstringspaces=true,
}

\lstdefinestyle{LLVMStyle}{
  numbers=none,
  stepnumber=0,
  numbersep=10pt,
  tabsize=4,
  showspaces=false,
  showstringspaces=true,
}

\lstdefinestyle{CStyle}{
  numbers=none,
  stepnumber=1,
  numbersep=10pt,
  tabsize=4,
  showspaces=false,
  showstringspaces=true,
}
\lstdefinelanguage{WAT}{
    otherkeywords={},
    morekeywords=[1]{i32,f32,i64,f64},
    morekeywords=[2]{0},
    morekeywords=[3]{add,const,mul,shl,get,rem_s,rem_u,ne,tee,sub,set,store},
    morekeywords=[4]{},
    morekeywords=[5]{global, get_global, mut, set_global, export, import,loop, memory, data, get_local,if, block,module, set_local,call,br_if,end, all,call_indirect,local,global,module, func, param, result, type},
    morekeywords=[6]{=,;},
    morekeywords=[7]{(,),[,],.},
    sensitive=false,
    morecomment=[l]{;},
    morecomment=[s]{;}{;},
    morestring=[b]",
    keywordstyle=[1]\color{eminence}\bfseries,
    keywordstyle=[3]\color{frenchplum},
    keywordstyle=[5]\color{darkgreen}\bfseries,
    commentstyle=\color{commentgreen}
}
\makeatletter
\lstdefinelanguage{CURL}{
    otherkeywords={},
    morekeywords=[1]{\<, <},
    morekeywords=[2]{0},
    morekeywords=[3]{<},
    morekeywords=[4]{},
    morekeywords=[5]{xpath, xtime, xpop, xpophash, date},
    morekeywords=[6]{=,;},
    morekeywords=[7]{(,),[,],.},
    sensitive=false,
    morecomment=[l]{;},
    morecomment=[s]{;}{;},
    morestring=[b]",
    keywordstyle=[1]\color{eminence}\bfseries,
    keywordstyle=[3]\color{frenchplum},
    keywordstyle=[5]\color{darkgreen}\bfseries,
    commentstyle=\color{commentgreen}
}
\lstdefinelanguage{pseudo}{
    otherkeywords={},
    morekeywords=[1]{\{, \}},
    morekeywords=[2]{0},
    morekeywords=[3]{<},
    morekeywords=[4]{switch, return, call},
    morekeywords=[5]{case},
    morekeywords=[6]{=,;},
    morekeywords=[7]{(,),[,],.},
    sensitive=false,
    morestring=[b]",
    keywordstyle=[1]\color{eminence}\bfseries,
    keywordstyle=[3]\color{frenchplum},
    keywordstyle=[5]\color{darkgreen}\bfseries,
    commentstyle=\color{commentgreen}
}
\lstdefinelanguage{llvm}{
    morecomment = [l]{;},
    morestring=[b]", 
    sensitive = true,
    morekeywords=[2]{i32,f32,i64,f64},
    morekeywords=[3]{
        define, declare, global, constant,
        internal, external, private,
        linkonce, linkonce_odr, weak, weak_odr, appending,
        common, extern_weak,
        thread_local, dllimport, dllexport,
        hidden, protected, default,
        except, deplibs,
        volatile, fastcc, coldcc, cc, ccc,
        x86_stdcallcc, x86_fastcallcc,
        ptx_kernel, ptx_device,
        signext, zeroext, inreg, sret, nounwind, noreturn,
        nocapture, byval, nest, readnone, readonly, noalias, uwtable,
        inlinehint, noinline, alwaysinline, optsize, ssp, sspreq,
        noredzone, noimplicitfloat, naked, alignstack,
        module, asm, align, tail, to,
        addrspace, section, alias, sideeffect, c, gc,
        target, datalayout, triple,
        blockaddress
    },
    morekeywords=[4]{
        fadd, sub, fsub, mul, fmul,
        sdiv, udiv, fdiv, srem, urem, frem,
        and, or, xor,
        icmp, fcmp,
        eq, ne, ugt, uge, ult, ule, sgt, sge, slt, sle,
        oeq, ogt, oge, olt, ole, one, ord, ueq, ugt, uge,
        ult, ule, une, uno,
        nuw, nsw, exact, inbounds,
        phi, call, select, shl, lshr, ashr, va_arg,
        trunc, zext, sext,
        fptrunc, fpext, fptoui, fptosi, uitofp, sitofp,
        ptrtoint, inttoptr, bitcast,
        ret, br, indirectbr, switch, invoke, unwind, unreachable,
        malloc, alloca, free, load, store, getelementptr,
        extractelement, insertelement, shufflevector,
        extractvalue, insertvalue,
    },
    alsoletter={\%},
    keywordsprefix={\%},% All identifiers starting with '%' will be printed as first order keywords.
    keywordstyle=[1]\bfseries,% As mentioned above, these are the keywords starting with '%', like '%5'
    keywordstyle=[2]\color{eminence}\bfseries,
    keywordstyle=[3]\color{darkgreen}\bfseries,
    keywordstyle=[4]\color{frenchplum},
}
\newenvironment{btHighlight}[1][]
{\begingroup\tikzset{bt@Highlight@par/.style={#1}}\begin{lrbox}{\@tempboxa}}
{\end{lrbox}\bt@HL@box[bt@Highlight@par]{\@tempboxa}\endgroup}

\newcommand\btHL[1][]{%
  \begin{btHighlight}[#1]\bgroup\aftergroup\bt@HL@endenv%
}
\def\bt@HL@endenv{%
  \end{btHighlight}%   
  \egroup
}
\newcommand{\bt@HL@box}[2][]{%
  \tikz[#1]{%
    \pgfpathrectangle{\pgfpoint{1pt}{0pt}}{\pgfpoint{\wd #2}{\ht #2}}%
    \pgfusepath{use as bounding box}%
    \node[anchor=base west, fill=orange!30,outer sep=0pt,inner xsep=1pt, inner ysep=0pt, rounded corners=3pt, minimum height=\ht\strutbox+1pt,#1]{\raisebox{1pt}{\strut}\strut\usebox{#2}};
  }%
}

\makeatother

\newtheorem{metric}{\textbf{Metric}}

\newtheorem{definition}{Definition}

\AtBeginDocument{%
  \providecommand\BibTeX{{%
    \normalfont B\kern-0.5em{\scshape i\kern-0.25em b}\kern-0.8em\TeX}}}

\title{Multi-variant Execution at the Edge}

\author{Javier Cabrera-Arteaga}
\email{javierca@kth.se}
\affiliation{%
  \institution{KTH Royal Institute of technology}
  \country{Sweden}
}

\author{Pierre Laperdrix}
\email{pierre.laperdrix@inria.fr}
\affiliation{%
  \institution{CNRS}
  \country{France}
}

\author{Martin Monperrus}
\email{martin.monperrus@kth.se}
\affiliation{%
  \institution{KTH Royal Institute of Technology}
  \country{Sweden}
}

\author{Benoit Baudry}
\email{baudry@kth.se}
\affiliation{%
  \institution{KTH Royal Institute of Technology}
  \country{Sweden}
}

\newcommand{\tool}{MEWE\xspace}

\newcommand{\projectcount}{2\xspace}

\newcommand{\ecount}{ 7\xspace }

\newcommand{\blue}[1]{{#1}}
\newcommand{\etal}{et al.\@\xspace}

\newcommand{\wasm}{WebAssembly\xspace}

\newlistof{todo}{td}{List of TODOs}

\begin{document}

\begin{abstract}

    Edge-Cloud computing offloads parts of the computations that traditionally occurs in the cloud to edge nodes. The binary format \wasm is increasingly used to distribute and deploy services on such platforms. Edge-Cloud computing providers let their clients deploy stateless services in the form of \wasm binaries, which are then translated to machine code, sandboxed and executed at the edge. 
    In this context, we propose a technique that (i) automatically diversifies \wasm binaries that are deployed to the edge and (ii) randomizes execution paths at runtime. Thus, an attacker cannot exploit all edge nodes with the same payload. Given a service, we automatically synthesize functionally equivalent variants for the functions providing the service. All the variants are then wrapped into a single multivariant \wasm binary. When the service endpoint is executed, every time a function is invoked, one of its variants is randomly selected. We implement this technique in the \tool tool and we validate it with 7 services for which \tool generates multivariant binaries that embed hundreds of function variants. We execute the multivariant binaries on the world-wide edge platform provided by Fastly, as part as a research collaboration. We show that multivariant binaries exhibit a real diversity of execution traces across the whole edge platform distributed around the globe.
    \end{abstract}


\maketitle

% \keywords{Edge, Cloud, WebAssembly}

\section{Introduction}

Edge-Cloud computing distributes a part of the  data and computation to edge nodes \cite{choy2014hybrid,taleb2017multi}.
Edge nodes are servers located in many countries and regions so that Internet resources get closer to the end users, in order to reduce latency and save bandwidth.
Video and music streaming services, mobile games, as well as e-commerce and news sites leverage this new type of cloud architecture to increase the quality of their services. 
For example, the New York Times website was able to serve more than 2 million concurrent visitors during the 2016 US presidential election with no difficulty thanks to Edge computing~\cite{fastlyNYT}.

The state of the art of edge computing platforms like Cloudflare or Fastly use the binary format \wasm (aka Wasm) \cite{CloudflareWasm, FastlyWasm} to deploy and execute on edge nodes. 
\wasm is a portable bytecode format designed to be lightweight, fast and safe \cite{haas2017bringing,bryant2020webassembly}.
After compiling code to a \wasm binary, developers spawn an edge-enabled compute service by deploying the binary on all nodes in an Edge platform.
Thanks to its simple memory and computation model, \wasm is considered safe~\cite{pMendkiServerless}, yet
is not exempt of vulnerabilities either at the execution engine's level \cite{ChromeZero} or the binary itself ~\cite{usenixWasm2020}.
Implementations in both, browsers and standalone runtimes~\cite{Narayan2021Swivel}, have been found to be vulnerable~\cite{usenixWasm2020,Narayan2021Swivel}.
This means that if one node in an Edge network is vulnerable, all the others are vulnerable in the exact same manner. 
In other words, the same attacker payload would break all edge nodes at once \cite{o2004achieving}.
This illustrates how Edge computing is fragile with respect to systemic vulnerabilities for the whole network, like it happened on June 8, 2021 for Fastly \cite{BREAKFastly}.

In this work, we introduce Multivariant Execution for WebAssembly in the Edge (\tool), a framework that generates diversified \wasm binaries so that no two executions in the edge network are identical.
Our solution is inspired by N-variant systems \cite{cox06}  where diverse variants are assembled for secretless security. 
Here, our goal is to drastically increase the effort for exploitation through large-scale execution path randomization.
\tool operates in two distinct steps. 
At compile time, \tool generates \textit{variants} for different functions in the program. 
A function variant  is semantically identical to the original function but structurally different, i.e., binary instructions are in different orders or have been replaced with equivalent ones.
All the function variants for one service are then embedded in a single multivariant \wasm binary. 
At runtime, every time a function is invoked, one of its variant is randomly selected. This way, the actual execution path taken to provide the service is randomized each time the service is executed hardening break-once-break-everywhere attacks.

We experiment \tool with 7 services, composed of hundreds of functions. 
We successfully synthesize thousands of function variants, which create orders of magnitude more possible execution paths than in the original service. 
To determine the runtime randomness of the embedded paths, we deploy and run the mutlivariant binaries on the Fastly edge computing platform (leading CDN platform). 
We collaborated with Fastly to experiment \tool on the actual production edge computing nodes that they provide to their clients.
This means that all our experiments ran in a real-world setting.
For this experiment, we execute each multivariant binary several times on every edge computing node provided by Fastly. 
Our experiment shows that the multivariant binaries render the same service as the original, yet with highly diverse execution traces. 

% Novelty statement
The novelty of our contribution is as follows.
First, we are the first to perform software diversification in the context of edge computing, with experiments performed on a real-world, large-scale, commercial edge computing platform (Fastly).
Second, very few works have looked at software diversity for \wasm \cite{CabreraArteaga2020CROWCD,Narayan2021Swivel}, our paper contributes to proving the feasibility of this challenging endeavour.

To sum up, our contributions are:

\begin{itemize}
    \item \tool: a framework that builds multivariant \wasm  binaries for edge computing, combining the automatic synthesis of  semantically equivalent function variants, with  execution path randomization. 
    \item Original results on the large-scale diversification of \wasm binaries, at the function and execution path levels.
    \item Empirical evidence of the feasibility of deploying our novel multivariant execution scheme on a real-world edge-computing platform.
    \item A publicly available prototype system, shared for future research on the topic:  \repourl.
\end{itemize}

This work is structured as follows. 
First, Section~\ref{sec:background} present a background on \wasm and its usage in an edge-cloud computing scenario. 
Section~\ref{sec:mewe} introduces the architecture and foundation of \tool while Section~\ref{sec:experiments} and Section~\ref{sec:results} present the different experiments we conducted to show the feasibility of our approach. Section~\ref{sec:related} details the Related Work while %Section~\ref{sec:discussion} discusses our results while
Section~\ref{sec:conclusion} concludes this paper.

\section{Background}
\label{sec:background}

In this section we introduce \wasm, as well as the deployment model that edge-cloud platforms such as Fastly provide to their clients. This forms the technical context for our work.

\subsection{\wasm}

\wasm is a bytecode designed to bring safe, fast, portable and compact low-level code on the Web.
The language was first publicly announced in 2015 and formalized by Haas \etal~\cite{haas2017bringing}. Since then, most major web browsers have implemented support for the standard.
Besides the Web, \wasm is independent of any specific hardware, which means that it can run in standalone mode. 
This allows for the adoption of \wasm outside web browsers \cite{bryant2020webassembly}, e.g., for  edge computing \cite{Narayan2021Swivel}.

\wasm binaries are usually compiled from source code like C/C++ or Rust. Listing \ref{CExample}  and \ref{WasmExample}  illustrate an example of a C function turned into \wasm.
Listing \ref{CExample} presents the C code of one function and \autoref{WasmExample} shows the result of compiling this function into a \wasm module. The \wasm code is further interpreted or compiled ahead of time into machine code.

\begin{code}
\lstset{language=C,caption={C function that calculates the quantity $2x + x$},label=CExample, frame=b}
\begin{lstlisting}[style=CStyle]
int f(int x) { 
    return 2 * x + x; 
}
\end{lstlisting}
\lstset{
    language=WAT,
    caption={\wasm code  for \autoref{CExample}.},
    style=WATStyle,
    stepnumber=0,
    frame=b,
    label=WasmExample}
\begin{lstlisting}
(module
  (type (;0;) (func (param i32) (result i32)))
  (func (;0;) (type 0) (param i32) (result i32)
    local.get 0
    local.get 0
    i32.const 2
    i32.mul
    i32.add)
    (export "f" (func 0)))
\end{lstlisting}
\vspace{-9mm}
\end{code}

\subsection{Web Assembly and Edge Computing}
\label{edge}

Using Wasm as an intermediate layer is better in terms of startup and memory usage, than containerization or virtualization \cite{pMendkiServerless, 1244493Jacobsson}. 
This has encouraged edge computing platforms like Cloudflare or Fastly to adopt \wasm to deploy client applications in a modular and sandboxed manner  \cite{CloudflareWasm, FastlyWasm}.
In addition, \wasm is a compact representation of code, which saves bandwidth when transporting code over the network .
%This allows edge-cloud platform providers to deploy the same Wasm binary around the world in a few seconds.

%\begin{figure}[!htbp]
%  \includegraphics[width=\linewidth]{diagrams/edge.pdf}
%  \caption{Deployment to the edge. Client application developers build their application and compile it to \wasm before submitting to the Edge-Cloud computing platform. Then, the Wasm binary is distributed to all edge nodes,  where it is compiled to a sandboxed machine code (x86 or ARM). This machine code is executed at every service request. }
%  \label{edge_upload_compilation}
%\end{figure}

Client applications that are designed to be deployed on edge-cloud computing platforms are usually isolated services, having one single responsibility. 
This development model is known as serverless computing, or function-as-a-service \cite{shillaker2020faasm,Narayan2021Swivel}. 
%\autoref{edge_upload_compilation} summarizes the development and deployment process for a client application to an edge-cloud platform.
The developers of a client application implement the isolated services in a given programming language. 
The source code and the HTTP harness of the service are then compiled to \wasm.
When client application developers deploy a \wasm binary, it is sent to all edge nodes in the platform. 
Then, the \wasm binary is compiled on each node to machine code. Each binary is compiled in a way that ensures that the code runs inside an isolated sandbox.

\subsection{Multivariant Execution}

In 2006, security researchers of University of Virginia have laid the foundations of a novel approach to security that consists in executing multiple variants of the same program. They called this ``N-variant systems'' \cite{cox06}. This potent idea has been renamed soon after as ``multivariant execution''. 

%\begin{definition}{Multivariant execution (MVE)}\label{def:mve}
%consists of creating multiple variants of a program and executing them in a principled way so as to improve security and reliability.
%\end{definition}

There is a wide range of realizations of MVE in different contexts.
% 2007 @inproceedings{bruschi2007diversified, title={Diversified process replic{\ae} for defeating memory error exploits},
Bruschi et al. \cite{bruschi2007diversified} and Salamat et al. \cite{salamat2007stopping} pioneered the idea of executing the variants in parallel.
% 2010 jackson2010multi Multi-variant program execution for vulnerability detection and analysis
% 2018 @article{lu2018stopping,  title={Stopping memory disclosures via diversification and replicated execution},
Subsequent techniques focus on MVE for mitigating memory vulnerabilities \cite{jackson2010multi,lu2018stopping}
% 2015 volckaert2015cloning  title={Cloning your gadgets: Complete ROP attack immunity with multi-variant execution
and other specific security problems incl. return-oriented programming attacks \cite{volckaert2015cloning} and code injection \cite{SalamatJWWF11}.
% 2019 @inproceedings{osterlund2019kmvx,  title={kMVX: Detecting kernel information leaks with multi-variant execution},
% salamat2009orchestra 
A key design decision of MVE is whether it is achieved in kernel space \cite{osterlund2019kmvx}, in user-space \cite{salamat2009orchestra},
% 2016 @inproceedings{koning2016secure, title={Secure and efficient multi-variant execution using hardware-assisted process virtualization},
with exploiting hardware features \cite{koning2016secure}, or even throught code polymorphism \cite{10.1145/3281662}.
Finally, one can neatly exploit the limit case of executing only two variants \cite{maurer2012tachyon,Kim2015}.
The body of research on MVE in a distributed setting has been less researched.
% 2021 voulimeneas2021dmvx dMVX: Secure and Efficient Multi-Variant Execution in a Distributed Setting
Notably, Voulimeneas \etal proposed a multivariant execution system by parallelizing the execution of the variants in different machines \cite{voulimeneas2021dmvx} for sake of efficiency.

In this paper, we propose an original kind of MVE in the context of edge computing. We generate multiple program variants, which we execute on edge computing nodes. We use the natural redundancy of Edge-Cloud computing architectures to deploy an internet-based MVE. Next section goes into the details of our procedure to generate variants and assemble them into multivariant binaries.

\section{\tool: Multivariant Execution for Edge Computing}
\label{sec:mewe}

In this section we present \tool, a novel technique to synthesize multivariant binaries and deploy them on an edge computing platform.

\subsection{Overview}

The goal of \tool is to synthesize multivariant  \wasm binaries, according to the threat model presented in \autoref{sec:threat-model}. 
The tool generates application-level multivariant binaries, without any change to the operating system or \wasm runtime.
The core idea of \tool is to synthesize diversified function variants providing execution-path randomization, according to the diversity model presented in \autoref{sec:diversity-model}.

\begin{figure*}
  \centering
  \includegraphics[width=0.7\linewidth]{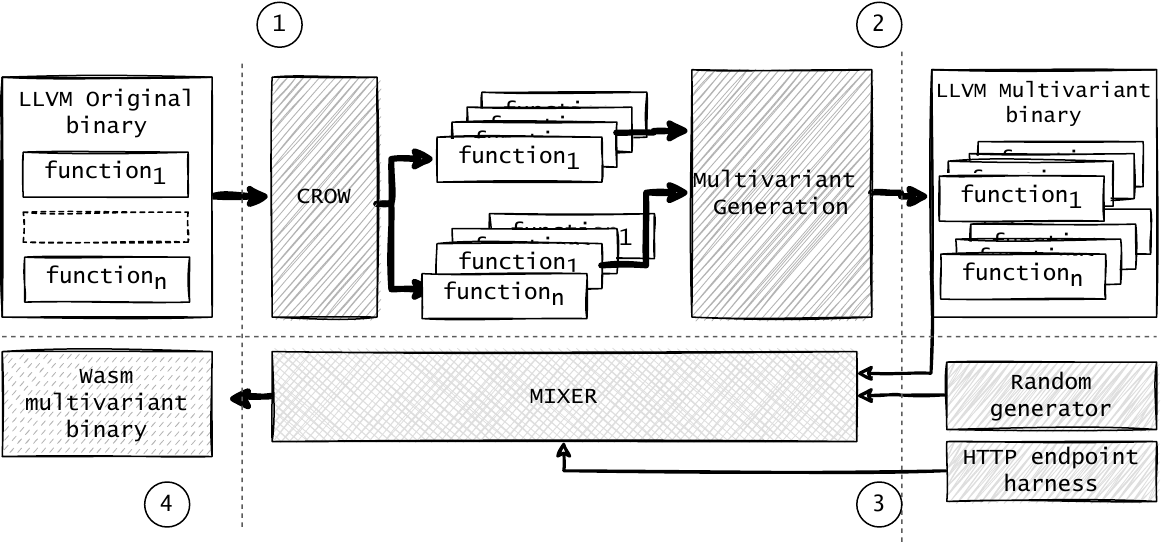}
  \caption{Overview of \tool. It takes as input the LLVM binary representation of a service composed of multiple functions. It first generates a set of functionally equivalent variants for each function in the binary and then generates a LLVM multivariant binary composed of all the function variants as well as dispatcher functions in charge of selecting a variant when a function is invoked. The \tool mixer composes the LLVM multivariant binary with a random number generation library and an edge specific HTTP harness, in order to produce a \wasm multivariant binary accessible through an HTTP endpoint and ready to be deployed to the edge. }
  \label{workflow}
\end{figure*}

In \autoref{workflow}, we summarize the analysis and transformation pipeline of \tool.
We pass a bitcode to be diversified, as an input to \tool.
Analysis and transformations are performed at the level of LLVM's intermediate representation (LLVM IR), as it is the best format for us to perform our modifications (see \autoref{sec:why-llvm}).
LLVM binaries can be obtained from any language with an LLVM frontend such as C/C++, Rust or Go, and they can easily be compiled to WebAssembly.
In Step~\step{1}, the binary is passed to CROW~\cite{CabreraArteaga2020CROWCD}, which is a superdiversifier for Wasm that generates a set of  variants for the functions in the binary. 
Step~\step{2} packages all the variants in one single multivariant LLVM binary. 
In Step~\step{3}, we use a special component, called a ``mixer'',  which augments the binary with two different components: an HTTP endpoint harness and a random generator, which are both required for executing Wasm at the edge. 
The harness is used to connect the program to its execution environment while the generator provides support for random execution path at runtime.
The final output of Step~\step{4} is a standalone multivariant \wasm binary that can be deployed on an edge-cloud computing platform. 
In the following sections, we describe in greater details the different stages of the workflow.

\subsection{Key design choices}
\label{sec:design}

In this section, we introduce the main design decisions behind \tool, starting from the threat model, to aligning the code analysis and transformation techniques.

\subsubsection{Threat Model}
\label{sec:threat-model}
With \tool, we aim to defend against an attacker that wants to leak private and sensitive information by learning about the state of the system and its characteristics.
These attacks include but are not limited to timing specific operations \cite{bernstein2005cache,aciiccmez2007cache}, counting register spill/reload operations to study and attack memory side-channels \cite{rane2015raccoon} and performing call stack analysis.
They can be performed either locally or remotely by finding a vulnerability or using shared resources in the case of a multi-tenant Edge computing server but the details of such exploitation are out of scope of this study.

%\subsubsection{Dangers related to Edge Computing}
%To benefit from the performance improvements offered by edge computing,  Fastly and Cloudflare modularize their services into a set of \wasm functions, which are deployed on all the edge nodes provided by the edge computing platforms. Then, these services are spawned on demand. This model of distributing the exact same \wasm binary on hundreds of computation nodes around the world is a serious risk for the infrastructure: a malicious developer who manages to exploit one vulnerability in the binary can exploit all the compute nodes with the same attack vector \cite{ }.

\subsubsection{Execution Diversification Model}
\label{sec:diversity-model}

\tool is designed to randomize the execution of \wasm programs, via diversification transformations.
Per Crane et al. those transformations are made to hinder side-channel attacks \cite{crane2015thwarting}. All programs are diversified with behavior preservation guarantees \cite{CabreraArteaga2020CROWCD}.
The core diversification strategies are:
\begin{enumerate*}
% https://en.wikipedia.org/wiki/Constant_folding
    \item \emph{Constant Inferring}. \tool identifies variables whose value can computed at compile time and are used to control branches. This has an effect on program execution times \cite{brennan2020jit}.
    \item \emph{Call Stack Randomization}. \tool introduces equivalent synthetic functions that are called randomly. This results in randomized call stacks, which complicates attacks based on call stack analysis \cite{liljestrand2021pacstack}. 
    
    \item \emph{Inline Expansion}. \tool inlines methods when appropriate. This also results in different call stacks, to hinder the same kind of attacks as for call stack randomization \cite{liljestrand2021pacstack}. 
    
    \item \emph{Spills/Reloads}. By performing semantically equivalent transformations for arithmetic expressions, the number of register spill/reload operations changes. Therefore, this changes the memory accesses in the machine code that is executed, affecting the measurement of memory side-channels \cite{rane2015raccoon}. 
\end{enumerate*}

\subsubsection{Diversification at the LLVM level}
\label{sec:why-llvm}

\tool diversifies programs at the LLVM level. Other solutions would have been to diversify at the source code level \cite{allier:hal-01089268}, or at the native binary level, eg x86 \cite{coppens2013feedback}. However, the former would limit the applicability of our work. The latter is not compatible with edge computing: the top edge computing execution platforms, e.g. Cloudflare and Fastly, mostly take \wasm binaries as input. 

LLVM, on the contrary, does not suffer from those limitations: 1) it supports different languages, with a rich ecosystem of frontends 2) it can reliably be retargeted to \wasm, thanks to the corresponding mature component in the LLVM toolchain.
%In addition, the LLVM ecosystem as a whole is very active, providing us with many different tools to facilitate our research endeavour.

\subsection{Variant generation}

\blue{
% How CROW works ?
\tool relies on the superdiversifier CROW \cite{CabreraArteaga2020CROWCD} to automatically diversify each function in the input LLVM binary (Step~\step{1}). CROW receives an LLVM module, analyzes the binary at the function block level and generates semantically equivalent variants for each function, if they exist. 
A function variant for \tool is semantically equivalent to the original (i.e., same input/output behavior), but exhibits a different internal behavior through tracing. 
Since the variants created by CROW are artificially synthesized from the original binary, after Step \step{1}, they are necessarily equivalent to the original program.
}
\blue{

}

\subsection{Combining variants into multivariant functions}

Step~\step{2} of \tool consists in combining the variants generated for the original functions, into into a single binary.
The goal is to support execution-path randomization at runtime.
% General idea
The core idea is to introduce one dispatcher function per original function for which we generate variants.
A dispatcher function is a synthetic function that is in charge of choosing a variant at random, every time the original function is invoked during the execution.
The random invocation of different variants at runtime is a known randomization technique, for example used by Lettner et al. with sanitizers \cite{lettner2018partisan}.

With the introduction of dispatcher function,  \tool turns the original call graph into a multivariant call graph, defined as follows. 

\begin{definition}{Multivariant Call Graph (MCG):}\label{def:EP}
    A multivariant call graph is a call graph $\langle N,E \rangle$ where the nodes in $N$ represent all the functions in the binary and an edge $(f_1,f_2) \in E$ represents a possible invocation of $f_2$ by $f_1$  \cite{ryder1979}, where the nodes are typed. The nodes in $N$ have three possible types: a function present in the original program,  a generated function variant, or a dispatcher function.
\end{definition}

\begin{figure}
    \centering
  \includegraphics[width=.9\linewidth]{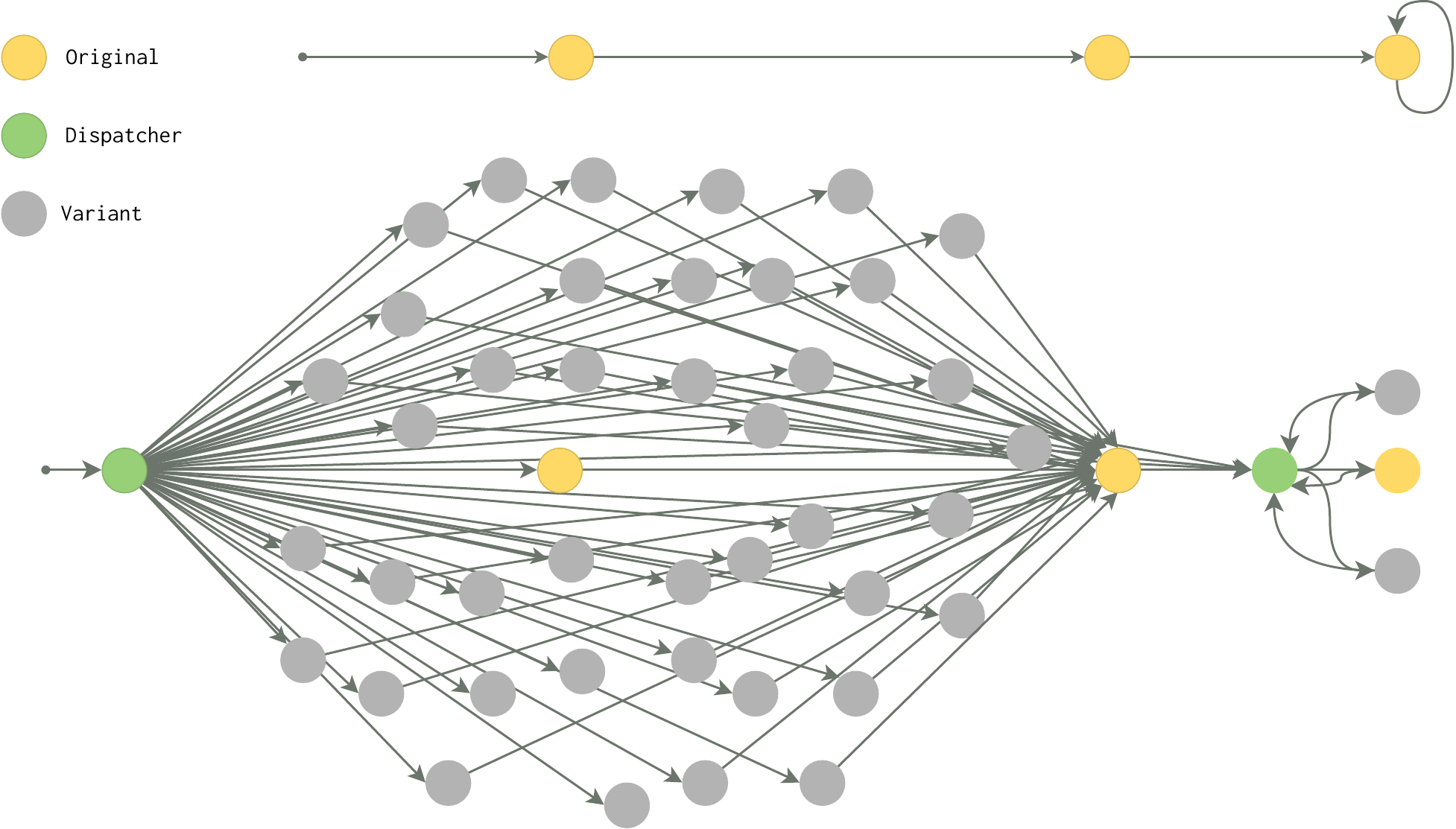}
  \caption{Example of two static call graphs for the bin2base64 endpoint of libsodium. At the top, the original call graph, at the bottom, the multivariant call graph, which includes nodes that represent function variants (in grey), dispatchers (in green), and original functions  (in yellow).
}
  \label{multivariant}
\vspace{-5mm}
\end{figure}

% Instance of a multivariant module
In \autoref{multivariant}, we show the original static call graph for program bin2base64 (top of the figure), as well as the multivariant call graph generated with \tool (bottom of the figure).
The grey nodes represent function variants, the green nodes function dispatchers and the yellow nodes are the original functions.
The possible calls are represented by the directed edges.
The original bin2base64 includes 3 functions. \tool generates 43 variants for the first function, none for the second and three for the third function. 
\tool introduces two dispatcher nodes, for the first and third functions. Each dispatcher is connected to the corresponding function variants, in order to invoke one variant randomly at runtime.

% exaplanation of dispatcher
The right most green node of \autoref{multivariant} is a function constructed as follows (See code in \autoref{dispatcher_code}).
% General logic of a multivartiant function
The function body first calls the random generator, which returns a value that is then used to invoke a specific function variant. 
It should be noted that the dispatcher function is constructed using the same signature as the original function. 

% Why a linear based switch
We implement the dispatchers with a switch-case structure to avoid indirect calls that can be susceptible to speculative execution based attacks \cite{Narayan2021Swivel}. 
The choice of a switch-case also avoids having multiple function definitions with the same signature, which could increase the attack surface in case the function signature is vulnerable \cite{johnson2021}.
This also allows \tool to inline function variants inside the dispatcher, instead of defining them again.
Here we trade security over performance, since dispatcher functions  that perform indirect calls, instead of a switch-case,  could improve the performance  of the dispatchers as indirect calls have constant time.

\subsection{\tool's Mixer}

The \tool mixer has four specific objectives: wrap functions as HTTP endpoints, link the LLVM multivariant binary, inject a random generator and merge all these components into a multivariant \wasm binary.

% Implementation
We use the Rustc compiler\footnote{\url{https://doc.rust-lang.org/rustc/what-is-rustc.html}} to orchestrate the mixing.% because Rustc is a compiler able to merge custom Rust source code with arbitratry-compatible LLVM binary producing a final Wasm binary.
% The random number generation
For the generator, we rely on WASI's specification \cite{WASI} for the random behavior of the dispatchers. Its exact implementation is dependent on the platform on which the binary is deployed.
% How to turn function into endpoints
For the HTTP harnesses, since our edge computing use case is based on the Fastly infrastructure, we rely on the Fastly API\footnote{\url{https://docs.rs/crate/fastly/0.7.3}} to transform our Wasm binaries into HTTP endpoints. The harness enables a function to be called as an HTTP request and to return a HTTP response.
Throughout this paper, we refer to an endpoint as the closure of invoked functions when the entry point of the \wasm binary is executed.

\subsection{Implementation}

The multivariant combination (Step \step{2}) is implemented in 942 lines of C++ code.
Its uses the LLVM 12.0.0 libraries to extend the LLVM standard linker tool capability with the multivariant generation.
\tool's Mixer (Step \step{3}) is implemented as an orchestration of the rustc and the \wasm backend provided by CROW.
An instantiation of how the multivariant binary works can be appreciated at \autoref{app:instance}.
For sake of open science and for fostering research on this important topic, the code of \tool is made publicly available on GitHub: \repourl.

%%%%%%%%%%%%%%%%%%%%%%%%%%%%%%%%%%%%%%%%%%%%%%%%%
%%%%%%%%%%%%%%% NEW SECTION %%%%%%%%%%%%%%%%%%%%%
%%%%%%%%%%%%%%%%%%%%%%%%%%%%%%%%%%%%%%%%%%%%%%%%%

\section{Experimental methodology}
\label{sec:experiments}

In this section we introduce our methodology to evaluate \tool. First, we present our research questions and the services with which we experiment the generation and the execution of multivariant binaries. Then, we detail the methodology for each research question.

\subsection{Research questions}

To evaluate the capabilities of \tool, we formulate the following research questions:

{%
% https://github.com/programming-journal/programming/issues/39
\setlength\leftmargini{2.8em}% Seems to be the original default. Fixes misalignment with custom label.
\begin{enumerate}[label=\textbf{RQ\arabic*}:, ref=RQ\arabic*]

\item \label{rq:multivariant} \textbf{(Multivariant Generation) How much diversity can \tool synthesize and embed in a multivariant binary ?}
\tool packages function variants in multivariant binaries. With this first question, we aim at measuring the amount of diversity that \tool can synthesize in the call graph of a program.

\item \label{rq:intramtd} \textbf{(Intra MVE) To what extent does \tool achieve multivariant executions on an edge compute node?} With this question we assess the ability of \tool to produce binaries that actually exhibit random execution paths when executed on one edge node. 

\item \label{rq:intermtd}\textbf{(Internet MVE) To what extent does \tool achieve multivariant execution over the worldwide Fastly infrastructure?} We check the diversity of execution traces gathered from the execution of a multivariant binary. The traces are collected from all edge nodes in order to assess MVE at a worldwide scale.

{
\blue{}

    \item \label{rq:performance} \textbf{What is the impact of the proposed multi-version execution on timing side-channels?} \tool generates binaries that embed a multivariant behavior. We measure to what extent \tool generates different execution times on the edge. Then, we discuss how multivariant binaries contribute to less predictable timing side-channels.
}

\end{enumerate}
 }

The core of the validation methodology for our tool \tool, consists in building multivariant binaries for several, relevant endpoints and to deploy and execute them on the Fastly edge-cloud platform.

\subsection{Study subjects}
\label{subjects}

We select two mature and typical edge-cloud computing projects to study the feasibility of \tool.
The projects are selected based on: suitability for  diversity synthesis with CROW (the projects should have the ability to collect their modules in LLVM intermediate representation), suitability for deployment on the Fastly infrastructure (the project should be easily portable Wasm/WASI and compatible with the Rust Fastly API), low chances to hit execution paths with no dispatchers and possibility to collect their execution runtime information (the endpoints should execute in a reasonable time of maximum 1 second even with the overhead of instrumentation). 
The selected projects are: \textbf{libsodium}, an encryption, decryption, signature and password hashing library which can be ported to WebAssembly and \textbf{qrcode-rust}, 
a QrCode and MicroQrCode generator written in Rust.

\definecolor{gray}{rgb}{0.9, 0.9, 0.9}
{
\begin{table}[h]
\vspace{-2mm}
\small
\centering
%\resizebox{\linewidth}{!}{
\begin{tabular}{  p{3.0cm} c c c }
    Name  & \#Endpoints & \#Functions & \#Instr. \\
    \hline    \hline

    \textbf{libsodium} & 5  & 62 & 6187 \\
    \url{https://github.com/jedisct1/libsodium} \\
    \hline
    \textbf{qrcode-rust} & 2  & 1840 &  127700 \\
    \url{https://github.com/kennytm/qrcode-rust} \\
    
\end{tabular}
%}

\caption{Selected projects to evaluate \tool: project name; the number of endpoints in the project that we consider for our experiments, the total number of functions to implement the endpoints, and the total number of \wasm instructions in the original binaries.
%This metadata is extracted from the Wasm binaries before they are deployed at the edge, thus, the number of functions might be different than in the source code
}\label{table:projects}
\vspace{-9mm}
\end{table}
}

In \autoref{table:projects}, we summarize some key metrics that capture the relevance of the selected projects.
The table shows the project name with its repository address, the number of selected endpoints for which we build multivariant binaries, the total number of functions included in the endpoints and the total number of Wasm instructions in the original binary.
Notice that, the metadata is extracted from the Wasm binaries before they are sent to the edge-cloud computing platform, thus, the number of functions might be not the same in the static analysis of the project source code

\subsection{Experiment's platform}

We run all our experiments on the Fastly edge computing platform. We deploy and execute the original and the multivariant endpoints on  64 edge nodes located around the world\footnote{The number of nodes provided in the whole platform is 72, we decided to keep only the 64 nodes that remained stable during our experimentation.}.
These edge nodes usually have an arbitrary and heterogeneous composition in terms of architecture and CPU model.
The deployment procedure is the same as the one described in \autoref{edge}.
The developers implement and compile their services to \wasm.
In the case of Fastly, the \wasm binaries need to be implemented with the Fastly platform API specification so they can properly deal with HTTP requests.
When the compiled binary is transmitted to Fastly, it is translated to x86 machine code with Lucet, which ensures the isolation of the service.

\subsection{RQ1 Multivariant diversity}
\label{method1}

We run \tool on each endpoint function of our \ecount endpoints.
In this experiment, we bound the search for function variant with timeout of 5 minutes per function. 
This produces one multivariant binary for each endpoint. 
To answer \ref{rq:multivariant}, we measure the number of function variants embedded in each  multivariant binary, as well as the number of execution paths that are added in the mutivariant call graphs, thanks to the function variants.

\subsection{RQ2 Intra MTD}
\label{method3}

We deploy the multivariant binaries of each of the \ecount endpoints presented in \autoref{table:CFG1}, on the 64 edge nodes of Fastly. 
We execute each endpoint, multiple times on each node, to measure the diversity of execution traces that are exhibited by the multivariant binaries. We have a time budget of 48 hours for this experiment. Within this timeframe,  we can query each endpoint 100 times on each node. Each query on the same endpoint is performed with the same input value. This is to guarantee that, if we observe different traces for different executions, it is due to the presence of multiple function variants. The input values are available as part of our reproduction package.

For each query, we collect the execution trace , i.e.,  the sequence of function names that have been executed when triggering the query.
To observe these  traces, we instrument the multivariant binaries to record each function entrance.

To answer RQ2, we measure the number of unique execution traces exhibited by each multivariant binary, on each separate edge node. To compare the traces, we hash them with the \texttt{sha256} function.
We then calculate the number of unique hashes among the 100 traces collected for an endpoint on one edge node.
We formulate the following definitions to construct the metric for RQ3.

\begin{metric}{Unique traces: $R(n, e)$.}\label{metric:ratio}
    Let $S(n, e)=\{T_1, T_2, ..., T_{100}\}$ be the collection of 100 traces collected for one endpoint $e$ on an edge node $n$, $H(n, e)$ the collection of hashes of each trace and $U(n, e)$ the set of unique trace hashes in $H(n,e)$.
    The uniqueness ratio of traces collected for edge node $n$ and endpoint $e$ is defined as
    $$
        R(n,e) = \frac{|U(n,e)|}{|H(n, e)|}
    $$
\end{metric}

The inputs that we pass to execute the endpoints at the edge and the received output for all executions are available in the reproduction repository at \repourl. 

\subsection{RQ3 Inter MTD}
\label{method4} We answer RQ3 by calculating the normalized Shannon entropy for all collected execution traces for each endpoint.
We define the following metric.

\begin{metric}{Normalized Shannon entropy: $E(e)$}\label{metric:entropy}
    Let $e$ be an endpoint, $C(e)=\cdot_{n=0}^{64} H(n, e)$ be the union  of all trace hashes for all edge nodes.
    The normalized Shannon Entropy for the endpoint $e$ over the collected traces is defined as: \\
    $$
        E(e)=-\Sigma \frac{p_x*log(p_x)}{log(|C(e)|)}
    $$
    Where $p_x$ is the discrete probability of the occurrence of the hash $x$ over $C(e)$.
    
\end{metric}

Notice that we normalize the standard definition of the Shannon Entropy by using the perfect case where all trace hashes are different. 
This normalization allows us to compare the calculated entropy between endpoints.
The value of the metric can go from 0 to 1. The worst entropy, value 0, means that the endpoint always perform the same path independently of the edge node and the number of times the trace is collected for the same node. On the contrary, 1 for the best entropy, when each edge node executes a different path every time the endpoint is requested.

The Shannon Entropy gives the uncertainty in the outcome of a sampling process.
If a specific trace has a high frequency of appearing in part of the sampling, then it is certain that this trace will appear in the other part of the sampling.

We calculate the metric for the \ecount endpoints, for 100 traces collected from 64 edge nodes, for a total of 6400 collected traces per endpoint.
Each trace is collected in a round robin strategy, i.e., the traces are collected from the 64 edge nodes sequentially.
For example, we collect the first trace from all nodes before continuing to the collection of the second trace.
This process is followed until 100 traces are collected from all edge nodes.

{

\blue{}

\subsection{RQ4 Timing side-channels}
\label{method5} 
For each endpoint listed in \autoref{table:CFG1}, we measure the impact of \tool on timing.
For this, we use the following metric:

\begin{metric}{Execution time:}\label{metric:time}
For a deployed binary on the edge, the execution time is the time spent on the edge to execute the binary.

%We refer to the backend-space as the time measured directly in the edge node, i.e., we instrument the code measure the time for the endpoint execution.
\end{metric}

Note that edge-computing platforms are, by definition, reached from the Internet.
Consequently, there may be latency in the timing measurement due to round-trip HTTP requests.
This can bias the distribution of measured execution times for the multivariant binary.
To avoid this bias, we instrument the code to only measure the execution on the  edge nodes.
}

We collect 100k execution times for each binary, both the original and multivariant binaries.
We perform a Mann-Withney U test \cite{mann1947} to compare both execution time distributions. 
If the P-value is lower than 0.05, two compared distributions are different. 
% https://support.minitab.com/en-us/minitab/19/help-and-how-to/statistics/nonparametrics/how-to/mann-whitney-test/interpret-the-results/all-statistics/

\section{Experimental Results}
\label{sec:results}

\subsection{RQ1 Results: Multivariant generation}

% Sum again the methodology
We use \tool to generate a multivariant binary for each  of the\ecount endpoints included in our \projectcount study subjects.  
We then calculate the number of diversified functions, in each endpoint, as well as how they combine to increase the number of  possible execution paths in the static call graph for the original and the multivariant binaries.

% Description of the table and general stats
The sections 'Original binary' and 'Multivariant WebAssembly binary' of
\autoref{table:CFG1} summarize the key data for RQ1. In the 'Original binary' section, the first column (\#F)  gives
the number of functions in the original binary and the second column (\#Paths) gives the number of possible execution paths in the original static call graph.
The 'Multivariant WebAssembly binary' section first shows the number of each type of nodes in the multivariant call graph: \#Non div. is the number of original functions that could not be diversified by \tool, \#D is the number of dispatcher nodes generated by \tool for each function that was successfully diversified, and \#V is the total number of function variants generated by \tool. The last column of this section is 
the number of possible execution paths in the static multivariant call graph.

\definecolor{celadon}{rgb}{0.67, 0.88, 0.69}
{
\begin{table}
\small
\centering
%\resizebox{\linewidth}{!}{
\begin{tabular}{  p{1.6cm} | r r | c  r r r }
     & \multicolumn{2}{c|}{\textbf{Original binary}} & \multicolumn{4}{c}{\textbf{Multivariant WebAssembly binary}} 
      \\
    \hline
    Endpoint & \#F  & \#Paths & \#Non D & \#D & \#V & \#Paths  \\
    \hline    \hline

    \textbf{libsodium} & & & & & & \\
    \hline
    
encrypt & 23 & 4 & 18 & 5 & 56  & 325  \\

decrypt & 20 & 3 & 16 & 5 & 49  & 84    \\

random & 8 & 2 & 6 & 2  & 238 & 12864  \\

invert & 8 & 2 & 6 & 2 & 125  & 2784  \\

bin2base64 & 3 & 2 & 1 & 2 & 47  & 172  \\

\hline
\textbf{qrcode-rust}  & & & & & & \\
\hline
qr\_str &  982 & 688$*10^6$ & 965 & 17 & 2092  & 97$*10^{12}$  \\
qr\_image & 858 &  1.4$*10^6$ & 843  & 15 & 2063  & 3$*10^{9}$   \\
\hline

\end{tabular}
%}

\caption{Static diversity generated by \tool, measured on the static call graphs of the \wasm binaries, and the  preservation of this diversity after translation to machine code. The table is structured as follows: Endpoint name; number of functions and numbers of possible paths in the original \wasm binary call graph; number of non diversified functions, number of created dispatchers (one per diversified functions), total number of function variants and number of execution paths in the multivariant \wasm binary call graph.}
%\footnote{
%For sake of simplicity, we use short names for the endpoints, the real names can be found in the reproduction repository.}

%}
\label{table:CFG1}
\vspace{-9mm}
\end{table}
}

% General stats saying that the CG is larger than 1, that means a lot of diversification
For all \ecount endpoints, \tool was able to diversify several functions and to combine them in order to increase the number of possible execution paths in several orders of magnitude.
For example, in the case of the \texttt{encrypt} function of libsodium, the original binary contains 23 functions that can be combined in 4 different paths. \tool generated a total of 56 variants for 5 of the 23 functions. These variants, combined with the 18 original functions in the multivariant call graph, form 325 execution paths. In other words, the number of possible ways to achieve the same encryption function has increased from 4 to 325, including dispatcher nodes that are in charge of randomizing the choice of variants at 5 different locations of the call graph. This increased number of possible paths, combined with random choices, made at runtime,  increases the effort a potential attacker needs to guess what variant is executed and hence what vulnerability she can exploit.

% Why a remarkable amount of execution paths
We have observed that there is no linear correlation between the number of diversified functions, the number of generated variants and the number of execution paths.
We have manually analyzed the endpoint with the largest number of possible execution paths in the multivariant Wasm binary: \texttt{qr\_str} of qrcode-rust.
\tool generated 2092 function variants  for this endpoint. 
Moreover, \tool inserted 17 dispatchers in the call graph of the endpoint. For each dispatcher, \tool includes between 428 and 3 variants.
If the original execution path contains function for which \tool is able to generate variants, then, there is a combinatorial explosion in the number of execution paths for the generated Wasm multivariant module.
The increase of the possible execution paths theoretically augments the uncertainty on which one to perform, in the latter case, approx. 140 000 times. 
As Cabrera and colleagues observed \cite{CabreraArteaga2020CROWCD} for CROW, a large presence of loops and arithmetic operations in the original function code leverages to more diversification.

Looking at the \#D (\#Dispatchers) and \#V (\#Variants) columns of the 'Multivariant WebAssembly binary' section of
\autoref{table:CFG1}, we notice that the number of variants generated per function greatly varies. For example, for both the \texttt{invert} and the \texttt{bin2base64} functions of Libsodium, \tool manages to diversify 2 functions (reflected by the presence of 2 dispatcher nodes in the multivariant call graph). Yet, \tool generates a total of 125 variants for the 2 functions in \texttt{invert}, and only 47 variants for the 2 functions in \texttt{bin2base64}. The main reason for this is related to the complexity of the diversified functions, which impacts the opportunities for the synthesis of code variations.

Columns \#Non D of the 'Multivariant WebAssembly binary' section of
\autoref{table:CFG1} indicates that, in each endpoint, there exists a number of functions for which  \tool did not manage to generate variants.
We identify three reasons for this, related to the diversification procedure of CROW, used by \tool to diversify individual functions. First, some functions cannot be diversified by CROW, e.g., functions that wrap only memory operations, which are oblivious to CROW diversification technique. 
Second, the complexity of the function directly affects the number of variants that CROW can generate.
Third, the diversification procedure of CROW is essentially a search procedure, which results are directly impacted by the tie budget for the search. In all experiments we give CROW 5 minutes maximum to synthesize function variants, which is a low budget for many functions.
It is important to notice that, the successful diversification of some functions in each endpoint, and their combination within the call graph of the endpoint, dramatically increases the number of possible paths that can triggered for multivariant executions.

\begin{tcolorbox}[boxrule=1pt,arc=.3em,boxsep=0mm]
    \textbf{Answer to RQ1}: \tool dramatically increases the  number  of possible execution paths in the multivariant \wasm binary of each endpoint.
    The large number of possible execution paths, combined with multiple points of random choice in the multivariant call graph thwart the prediction of which path will be taken at runtime.
\end{tcolorbox}

\subsection{RQ2 Results: Intra MTD}

To answer RQ2, we execute the multivariant binaries of each endpoint, on the Fastly edge-cloud infrastructure. 
We execute each endpoint 100 times on each of the 64 Fastly edge nodes.
All the executions of a given endpoint are performed with  the same input.
This allows us to determine if the execution traces are different due to the injected dispatchers and their random behavior.
After each execution of an endpoint, we collect the sequence of invoked functions, i.e., the execution trace. 
Our intuition is that the random dispatchers  combined with the function variants embedded in a multivariant binary are very likely to trigger different traces for the same execution, i.e., when an endpoint is executed several times in a row with the same input and on the same edge node.
The way both the function variants and the dispatchers contribute to exhibiting different execution traces is illustrated in \autoref{http:workflow}.

\begin{figure}
  \includegraphics[trim=900 600 600 900,clip,width=0.85\linewidth]{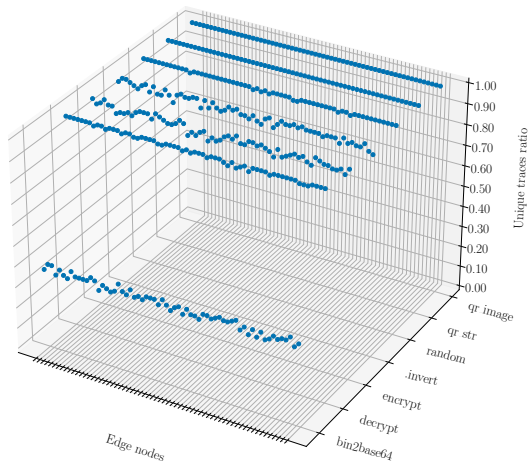}
  \caption{Ratio of unique execution traces for each endpoint on each edge node.   The X axis illustrates the edge nodes.
    The Y axis annotates the name of the endpoint.
    In the plot, for a given (x,y) pair, there is blue point representing the \autoref{metric:ratio} value in a set of 100 collected execution traces.}
  \label{diversity:traces}
\vspace{-7mm}
\end{figure}

\autoref{diversity:traces} shows the ratio of unique traces exhibited by each endpoint, on each of the 64 separate edge nodes. 
The X corresponds to the edge nodes.
The Y axis gives the name of the endpoint.
In the plot, for a given (x,y) pair, there is blue point in the Z axis representing  \autoref{metric:ratio} over 100 execution traces.

% large spread and non-zero in all cases
For all edge nodes, the ratio of unique traces is above 0.38.
In 6 out of 7 cases, we have observed that the ratio is remarkably high, above 0.9.
These results show that \tool  generates multivariant binaries that can randomize execution paths at runtime, in the context of an edge node. The randomization dispatchers, associated to a significant number of function variants greatly reduce the certainty about which computation is performed when running a specific input with a given input value.

Let's illustrate the phenomenon with the endpoint \texttt{invert}.
The endpoint \texttt{invert} receives a vector of integers and returns its inversion.
Passing a vector of integers with 100 elements as input, $I=[100,...,0]$, results in output $O=[0,...,100]$.
When the endpoint executes 100 times with the same input on the original binary, we observe 100 times the same execution trace.
When the endpoint is executed 100 times with the same input $I$ on the multivariant binary, we observe between 95 and 100 unique execution traces, depending on the edge node.
Analyzing the traces we observe that they include only two invocations to a dispatcher, one at the start of the trace and one at the end.
The remaining events in the trace are fixed each time the endpoint is executed with the same input $I$.
Thus, the maximum number of possible unique traces is the multiplication of the number of variants for each dispatcher, in this case $29\times96=2784$ . The probability of observing the same trace is $1/2784$.

For multivariant binaries that embed only a few variants, like in the case of the \texttt{bin2base64} endpoint, the ratio of unique traces per node is lower than for the other endpoints.
With the input we pass to \texttt{bin2base64}, the execution trace includes 57 function calls.
We have observed that, only one of these calls invokes a dispatcher, which can select among 41 variants. Thus,  probability of having the same execution trace  twice is $1/41$. 

Meanwhile, \texttt{qr\_str} embeds thousands of variants, and the input we pass triggers the invocation of 3M functions, for which 210666 random choices are taken relying on 17 dispatchers. Consequently, the probability of observing the same trace twice is infinitesimal. Indeed, all the executions of \texttt{qr\_str} are unique, on each separate edge node.
This is shown in \autoref{diversity:traces}, where the ratio of unique traces is 1 on all edge nodes. 

%On the other hand, Fastly claims to deploy a new version of the endpoint in 13s in average if it is request.
%This redeployment time makes unnecessary to test the collection of traces after 13 second.

\begin{tcolorbox}[boxrule=1pt,arc=.3em,boxsep=0.2mm]
    \textbf{Answer to RQ2}: Repeated executions of a multivariant binary with the same input on an individual edge node exhibits diverse execution traces.
    \tool successfully synthesizes multivariant binaries that trigger diverse execution paths at runtime, on individual edge nodes.
  
\end{tcolorbox}

\subsection{RQ3 Results: Internet MTD}
To answer RQ3, we build the union of all the execution traces  collected on all edge nodes for a given endpoint. 
Then, we compute the normalized Shannon Entropy over this set for each endpoint (\autoref{metric:entropy}).
Our goal is to determine whether the diversity of execution traces we observed on individual nodes in RQ3, actually generalizes to the whole edge-cloud infrastructure. Depending on many factors, such as the random number generator or a bug in the dispatcher, it could happen that we observe different traces on individual nodes, but that the set of traces is the same on all nodes. With RQ4 we assess the ability of \tool to exhibit multivariant execution at a global scale.

\autoref{table:entropy} provides the data to answer RQ3. The second column gives the normalized Shannon Entropy value (\autoref{metric:entropy}). Columns 3 and 4 give  the median and the standard deviation for the length of the execution traces. Columns 5 and 6 give 
the number of dispatchers that are invoked during the execution of the endpoint (\#ED) and the total number of invocations of these endpoints (\#Rch). These last two columns indicate to what extent the execution paths are actually randomized at runtime. In the cases of \texttt{invert} and \texttt{random}, both have the same number of taken random choices. However, the number of variants to chose in \texttt{random}  are larger, thus, the entropy, is larger than \texttt{invert}.

Overall, the normalized Shannon Entropy is above 42\%. This is evidence that the multivariant binaries generated by \tool can indeed exhibit a high degree of execution trace diversity, while keeping the same functionality. The number of randomization points along the execution paths (\#Rch) is at the core of these high entropy values. For example, every execution of the \texttt{encrypt} endpoint triggers 4M random choices among the different function variants embedded in the multivariant binaries. Such a high degree of randomization is essential to generate very diverse execution traces.

The \texttt{bin2base64} endpoint has the lowest level of diversity. As discussed in RQ2, this endpoint is the one that has the least variants and its execution path can be randomized only at one point. The low level of unique traces observed on individual nodes is reflected at the system wide scale with a globally low  entropy.

For both  \texttt{qr\_str} and \texttt{qr\_image} the entropy value is 1.0. This means that all the traces that we observe for all the executions of these endpoints are different from each other. In other words, someone who runs these services over and over with the same input cannot know exactly what code will be executed in the next execution.
These very high entropy values are made possible by the millions of random choices that are made along the execution paths of these endpoints.

% What is happening with the trace lengths
While there is a high degree of diversity among the traces exhibited by each endpoint, they all have the same length, except in the case of  \texttt{random}.
This means that the entropy is a direct consequence of the invocations of the dispatchers.
In the case of \texttt{random}, it naturally has a non-deterministic behavior.
Meanwhile, we observe several calls to dispatchers in during the execution of the multivariant binary, which indicates that \tool can amplify the natural diversity of traces exhibited by \texttt{random}. 
For each endpoint, we managed to trigger all dispatchers during its execution.
There is a correlation between the entropy and the number of random choices (Column \#RChs) taken during the execution of the endpoints.
For a high number of dispatchers, and therefore random choices, the entropy is large, like the cases of \texttt{qr\_str} and \texttt{qr\_image} show.
The contrary happens to \texttt{bin2base64} where its multivariant binary contains only one dispatcher.

\definecolor{celadon}{rgb}{0.67, 0.88, 0.69}
{
\begin{table}
\small
\centering
%\resizebox{\linewidth}{!}{
\begin{tabular}{  p{2.1cm} |  r | r r | r r }

    Endpoint & Entropy & MTL & $\sigma$ & \#ED & \#RCh \\
    \hline
    \hline
    
    \textbf{libsodium}  \\
    \hline
    
encrypt & 0.87 & 816 & 0 & 5  & 4M\\

decrypt & 0.96  & 440 & 0 & 5 & 2M\\

random & 0.98 & 15 & 5 & 2 & 12800\\

invert & 0.87  & 7343 & 0 & 2 & 12800\\

bin2base64 & 0.42  & 57 & 0 & 1 & 6400\\

\hline
\textbf{qrcode-rust} \\
\hline
qr\_str & 1.00 & 3045193 & 0 & 17 & 1348M \\
qr\_image & 1.00  & 3015450 & 0 & 15 & 1345M\\
\hline

\end{tabular}
%}
\caption{Execution trace diversity over the Fastly edge-cloud computing platform. The table is formed of 6 columns: the name of the endpoint, the normalized Shannon Entropy value (\autoref{metric:entropy}), the median size of the execution traces (MTL), the standard deviation for the trace lengths the number of executed dispatchers (\#ED) and the number of total random choices taken during all the 6400 executions (\#RCh).}\label{table:entropy}
\vspace{-8mm}
\end{table}
}

\begin{tcolorbox}[boxrule=1pt,arc=.3em,boxsep=0.2mm]
    \textbf{Answer to RQ3}: At the internet scale of the Edge platform, the multivariant binaries synthesized by \tool exhibit a massive diversity of execution traces, while still providing the original service.
    It is virtually impossible for an attacker to predict which is taken for a given query.
\end{tcolorbox}

{
\blue{}

\subsection{RQ4 Results: Timing side-channels} 
For each endpoint used in RQ1, we compare the execution time distributions for the original binary and the multivariant binary. All distributions are measured on 100k executions.
In \autoref{table:ex_times}, we show the execution time for the original endpoints and their corresponding multivariant. The table is structured in two sections. The first section shows the endpoint name, the median and standard deviation of the original endpoint. The second section shows the median and the standard deviation for the execution time of the corresponding multivariant binary.

\definecolor{celadon}{rgb}{0.67, 0.88, 0.69}
{
\begin{table}
\small
\centering
%\resizebox{\linewidth}{!}{
\begin{tabular}{  p{2.0cm}  r r | r r}
    & \multicolumn{2}{c|}{\textbf{Original bin.}} & \multicolumn{2}{c}{\textbf{Multivariant Wasm}} \\
    \hline
    Endpoint & Median ($\mu$s) & $\sigma$ & Median ($\mu$s) & $\sigma$ \\
    \hline
    \hline
    
    \textbf{libsodium}  \\
    \hline
    
encrypt  & 7 & 5 & 217 & 43 \\

decrypt  & 13 & 6 & 225 & 47 \\

random  & 16 & 7 & 232 & 53 \\

invert  & 119 & 34 & 341 & 65 \\

bin2base64  & 10 & 5 & 215 & 35 \\

\hline
\textbf{qrcode-rust} \\
\hline
qr\_str & 3,117 & 418 & 492,606 & 36,864 \\
qr\_image & 3,091 & 412 & 512,669 & 41,718 \\
\hline

\end{tabular}
%}
\caption{Execution time distributions for 100k executions, for the original endpoints and their corresponding multivariants. The table is structured in two sections. The first section shows the endpoint name, the median execution time and its standard deviation for the original endpoint. The second section shows the median execution time and its standard deviation for the multivariant \wasm binary.}\label{table:ex_times}
\vspace{-8.5mm}
\end{table}
}

% statistics
We also observe that the distributions for multivariant binaries  have a higher standard deviation of execution time.
A statistical comparison between the execution time distributions confirms the significance of this difference (P-value = 0.05 with a  Mann-Withney U test). This hints at the fact that  the execution time for multivariant binaries is more unpredictable than the time to execute the original binary. 

% Curve flatenning
In \autoref{diversity:times}, each subplot represents the distribution for a single endpoint, with the colors blue and green representing the original and multivariant binary respectively. These plots reveal  that the execution times are indeed spread over a larger range of values compared to the original binary. 
This is evidence that execution time is less predictable for multivariant binaries than for the original ones.

\begin{figure*}
    \centering
  \includegraphics[width=0.9\linewidth]{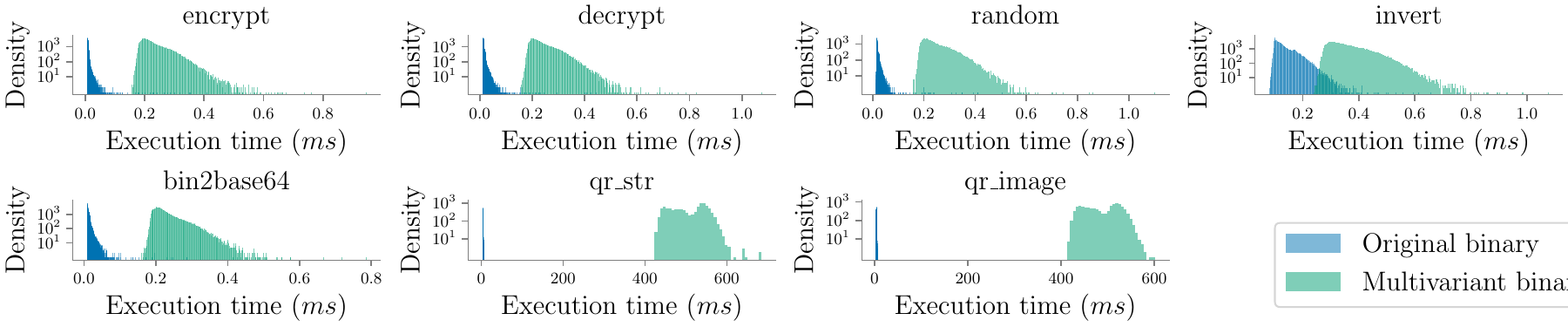}
  \caption{Execution time distributions. Each subplot represents the distribution for a single endpoint, blue for the original endpoint and green for the multivariant binary. The X axis shows the execution time in milliseconds and the Y axis shows the density distribution in logarithmic scale.}
  \label{diversity:times}
\vspace{-4mm}
\end{figure*}

%\subsubsection{Variant inferring}

We evaluate to what extent a specific variant can be detected by observing the execution time distribution. This evaluation is based on the measurement with one endpoint. For this, we choose endpoint \texttt{bin2base64} because it is the end point that has the least variants and the least dispatchers, which is the most conservative assumption. 
    
We dissect the collected execution times for the \texttt{bin2base64} endpoint, grouping them by execution path. In \autoref{diversity:distributions}, each opaque curve represents a cumulative execution time distribution of a unique execution path out of the 41 observed.
We observe that no specific distribution is remarkably different from another one. Thus, no specific variant can be inferred out of the projection of all execution times like the ones presented in \autoref{diversity:times}. Nevertheless, we calculate a Mann-Whitney test for each pair of distributions, $41\times41$ pairs. For all cases, there is no statistical evidence that the distributions are different, $P > 0.05$.

\begin{figure}
    \centering
  \includegraphics[trim=5 12 5 10,clip,width=0.7\linewidth]{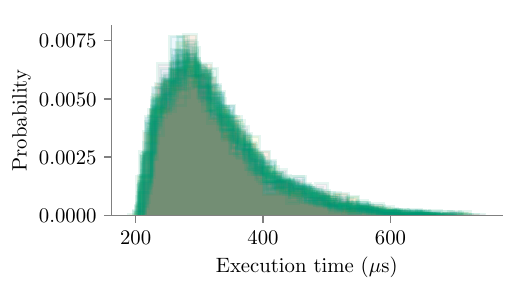}
  \caption{Execution time distributions for the \texttt{bin2base64} endpoint. Each opaque curve represents an execution time distribution of a unique execution path out of the 41 observed.}
  \label{diversity:distributions}
\vspace{-5mm}
\end{figure}

Recall that the choice of function variant is randomized at each function invocation, and the variants have different execution times as a consequence of the code transformations, i.e., some variants execute more instructions than others. 
Consequently, attacks relying on measuring precise execution times of a function are made a lot harder to conduct as the distribution for the multivariant binary is different and even more spread than the original one.

We note that the execution times are slower for multivariant binaries. Being under 500 ms in general, this does not represent a threat to the applicability of  multivariant execution at the edge. Yet, it calls for future optimization research.

%\vspace{-0.2cm}

\begin{tcolorbox}[boxrule=1pt,arc=.3em,boxsep=0mm]
    \textbf{Answer to RQ4}: The execution time distributions are significantly different between the original and the multivariant binary. Furthermore, no specific variant can be inferred from execution times gathered from the multivariant binary. \tool contributes to mitigate potential attacks based on predictable execution times.
\end{tcolorbox}

}

\section{Related Work}
\label{sec:related}

Our work is in the area of software  diversification for security, a research field discovered by researchers Forrest \cite{forrest97} and Cohen \cite{cohen93}. We contribute a novel technique for multivariant execution, and discuss related work in  \autoref{sec:background}. Here, we position our contribution with respect to previous work on randomization and security for \wasm.

\subsection{Related Work on Randomization}
\label{sec:rw-randomization}

A randomization technique creates a set of unique executions for the very same program \cite{bhatkar03}. Seminal works include instruction-set randomization \cite{Kc03,barrantes2003randomized}
to create a unique mapping between artificial CPU instructions and real ones. This makes  it very hard for an attacker ignoring the key to inject executable code. Compiler-based  techniques can randomly introduce NOP and padding to statically diversify programs. 
\cite{jackson} have explored how to use NOP and it breaks the predictability of program execution, even mitigating certain exploits to an extent.

Chew and Song \cite{Chew02mitigatingbuffer} target operating system randomization. They randomize the interface between the operating system and the user applications:
the system call numbers, the library entry points (memory addresses) and the stack placement. All those techniques are dynamic, done at runtime using load-time preprocessing and rewriting. 
Bathkar et al. \cite{bhatkar03,bhatkar2005efficient} have proposed  three kinds of randomization transformations: randomizing the base addresses of applications and libraries  memory regions, random permutation of the order of variables and routines, and the random introduction of random gaps between objects. 
Dynamic randomization can address different kinds of problems. In particular, it  mitigates a large range of memory error exploits. 
Recent work in this field include stack layout randomization against data-oriented programming \cite{aga2019smokestack} and memory safety violations \cite{lee2021savior}, as well as a technique to reduce the exposure time of persistent memory objects to increase the frequency of address randomization \cite{xu2020merr}.

We contribute to the field of randomization, at two stages. First, we automatically generate variants of a given program, which have different  \wasm code and still behave the same. Second, we randomly select which variant is executed at runtime, creating a multivariant execution scheme that randomizes the observable execution trace at each run of the program.

Davi \etal proposed Isomeron \cite{davi2015isomeron}, an approach 
for execution-path randomization.
Isomeron simultaneously loads the original program and a variant. While the program is running, Isomeron continuously flips a coin to decide which copy of the program should be executed next at the level of function calls. With this strategy, a potential attacker cannot predict whether the original or the variant of a program will execute.
\tool proposes two key novel contributions. First,  our code diversification step can generate variants of complex control flow structures by inferring constants or loop unrolling.
Second, \tool interconnects hundreds of variants and several randomization dispatchers in a single binary, increasing by orders of magnitude the runtime uncertainty about what code will actually run, compared to the choice among 2 variants proposed by Isomeron.
{
    \blue{}
%    On the same topic, Belleville \etal \cite{10.1145/3281662} proposed a polymorphism approach for multivariant execution. They created an LLVM-based compiler that generates programs for which random machine code is constructed at runtime. They showed the benefits of multivariant execution against side-channels attacks.  \tool performs the same behavior by interconnecting hundreds of variants in a single binary and allows a broader space of code transformations, such as constant inferring and loop unrolling.
}

\subsection{Related work on \wasm Security}
The reference piece about \wasm security is by Lehmann et al. \cite{usenixWasm2020}, which presents three attack primitives. Lehmann et al. have then followed up with a large-scale empirical study of \wasm binaries \cite{hilbig2021empirical}.
% WASM compiler Security
Narayan \etal \cite{Narayan2021Swivel} remark that the security model of \wasm is vulnerable to Spectre attacks.
This means that  \wasm sandboxes may be hijacked and leak memory.
They propose to modify the Lucet compiler used by Fastly to incorporate LLVM fence instructions\footnote{\url{https://llvm.org/doxygen/classllvm_1_1FenceInst.html}} in the machine code generation, trying to avoid speculative execution mistakes.
Johnson \etal \cite{johnson2021}, on the other hand, propose fault isolation for \wasm binaries, a technique that can be applied before being deployed to the edge-cloud platforms.
% Compositional Information Flow Analysis for WebAssembly Programs stievenart2020compositional
Stievenart et al. \cite{stievenart2020compositional} design a static analysis dedicated to information flow problems.
%   title={Minethrottle: Defending against wasm in-browser cryptojacking}, @inproceedings{bian2020minethrottle,
Bian et al. \cite{bian2020minethrottle} performs runtime monitoring of \wasm to detect cryptojacking.
The main difference with our work is that our defense mechanism is larger in scope, meant to tackle ``yet unknown'' vulnerabilities.
Notably, \tool is agnostic from the last-step compilation pass that translates Wasm to machine code, which means that the multivariant binaries can be deployed on any edge-cloud platform that can receive \wasm endpoints, regardless of the underlying hardware.

\section{Conclusion}
\label{sec:conclusion}

% Sumarize results
In this work we propose a novel technique to automatically synthesize multivariant binaries to be deployed on edge computing platforms. Our tool, \tool, operates on a single service implemented as a \wasm binary. It automatically generates functionally equivalent variants for each function that implements the service, and combines all the variants in a single \wasm binary, which exact execution path is randomized at runtime.
Our evaluation with \ecount real-world cryptography and QR encoding services shows that \tool can generate hundreds of function variants and combine them into binaries that include from thousands to millions of possible execution paths.
The deployment and execution of the multivariant binaries on the Fastly cloud platform  showed that they actually exhibit a very high diversity of execution at runtime, in single edge nodes, as well as Internet scale.
%Besides, \tool is able to sparse the execution times of multivariant \wasm binaries, encouraging it usage for timing-side channels hardening.

% Future work
Future work with \tool will address the trade-off between a large space for execution path randomization and the computation cost of large-scale runtime randomization. 
In addition, the synthesis of a large pool of variants supports the exploration of the concurrent execution of multiple variants to detect misbehaviors in services deployed at the edge.

%\IEEEtriggeratref{56} % Recommended way to balance columns is NOT with the balance package because it splits references in half
%\IEEEtriggeratref{} % Recommended way to balance columns is NOT with the balance package because it splits references in half

\bibliography{main}


\begin{thebibliography}{58}

%%% the Web address.

\ifx \showCODEN    \undefined \def \showCODEN     #1{\unskip}     \fi
\ifx \showDOI      \undefined \def \showDOI       #1{#1}\fi
\ifx \showISBNx    \undefined \def \showISBNx     #1{\unskip}     \fi
\ifx \showISBNxiii \undefined \def \showISBNxiii  #1{\unskip}     \fi
\ifx \showISSN     \undefined \def \showISSN      #1{\unskip}     \fi
\ifx \showLCCN     \undefined \def \showLCCN      #1{\unskip}     \fi
\ifx \shownote     \undefined \def \shownote      #1{#1}          \fi
\ifx \showarticletitle \undefined \def \showarticletitle #1{#1}   \fi
\ifx \showURL      \undefined \def \showURL       {\relax}        \fi
% invisible to TeX
\providecommand\bibfield[2]{#2}
\providecommand\bibinfo[2]{#2}
\providecommand\natexlab[1]{#1}
\providecommand\showeprint[2][]{arXiv:#2}

\bibitem[\protect\citeauthoryear{??}{BRE}{2021}]%
        {BREAKFastly}
 \bibinfo{year}{2021}\natexlab{}.
\newblock \bibinfo{title}{{Global CDN Disruption}}.
\newblock
\newblock
\urldef\tempurl%
\url{https://status.fastly.com/incidents/vpk0ssybt3bj}
\showURL{%
\tempurl}


\bibitem[\protect\citeauthoryear{??}{fas}{2021}]%
        {fastlyNYT}
 \bibinfo{year}{2021}\natexlab{}.
\newblock \bibinfo{title}{{The New York Times on failure, risk, and prepping
  for the 2016 US presidential election -- Fastly}}.
\newblock
\newblock
\urldef\tempurl%
\url{https://www.fastly.com/blog/new-york-times-on-failure-risk-and-prepping-2016-us-presidential-election}
\showURL{%
\tempurl}


\bibitem[\protect\citeauthoryear{??}{WAS}{2021}]%
        {WASI}
 \bibinfo{year}{2021}\natexlab{}.
\newblock \bibinfo{title}{WebAssembly System Interface}.
\newblock
\newblock
\urldef\tempurl%
\url{https://github.com/WebAssembly/WASI}
\showURL{%
\tempurl}


\bibitem[\protect\citeauthoryear{Ac{\i}i{\c{c}}mez, Schindler, and
  Ko{\c{c}}}{Ac{\i}i{\c{c}}mez et~al\mbox{.}}{2007}]%
        {aciiccmez2007cache}
\bibfield{author}{\bibinfo{person}{Onur Ac{\i}i{\c{c}}mez},
  \bibinfo{person}{Werner Schindler}, {and} \bibinfo{person}{{\c{C}}etin~K
  Ko{\c{c}}}.} \bibinfo{year}{2007}\natexlab{}.
\newblock \showarticletitle{Cache based remote timing attack on the AES}. In
  \bibinfo{booktitle}{\emph{Cryptographers’ track at the RSA conference}}.
  Springer, \bibinfo{pages}{271--286}.
\newblock


\bibitem[\protect\citeauthoryear{Aga and Austin}{Aga and Austin}{2019}]%
        {aga2019smokestack}
\bibfield{author}{\bibinfo{person}{Misiker~Tadesse Aga} {and}
  \bibinfo{person}{Todd Austin}.} \bibinfo{year}{2019}\natexlab{}.
\newblock \showarticletitle{Smokestack: thwarting DOP attacks with runtime
  stack layout randomization}. In \bibinfo{booktitle}{\emph{Proc. of CGO}}.
  \bibinfo{pages}{26--36}.
\newblock
\urldef\tempurl%
\url{https://drive.google.com/file/d/12TvsrgL8Wt6IMfe6ASUp8y69L-bCVao0/view}
\showURL{%
\tempurl}


\bibitem[\protect\citeauthoryear{Allier, Barais, Baudry, Bourcier, Daubert,
  Fleurey, Monperrus, Song, and Tricoire}{Allier et~al\mbox{.}}{2015}]%
        {allier:hal-01089268}
\bibfield{author}{\bibinfo{person}{Simon Allier}, \bibinfo{person}{Olivier
  Barais}, \bibinfo{person}{Benoit Baudry}, \bibinfo{person}{Johann Bourcier},
  \bibinfo{person}{Erwan Daubert}, \bibinfo{person}{Franck Fleurey},
  \bibinfo{person}{Martin Monperrus}, \bibinfo{person}{Hui Song}, {and}
  \bibinfo{person}{Maxime Tricoire}.} \bibinfo{year}{2015}\natexlab{}.
\newblock \showarticletitle{Multitier diversification in Web-based software
  applications}.
\newblock \bibinfo{journal}{\emph{{IEEE Software}}} \bibinfo{volume}{32},
  \bibinfo{number}{1} (\bibinfo{year}{2015}), \bibinfo{pages}{83--90}.
\newblock
\urldef\tempurl%
\url{https://doi.org/10.1109/MS.2014.150}
\showDOI{\tempurl}


\bibitem[\protect\citeauthoryear{Barrantes, Ackley, Forrest, Palmer,
  Stefanovic, and Zovi}{Barrantes et~al\mbox{.}}{2003}]%
        {barrantes2003randomized}
\bibfield{author}{\bibinfo{person}{Elena~Gabriela Barrantes},
  \bibinfo{person}{David~H Ackley}, \bibinfo{person}{Stephanie Forrest},
  \bibinfo{person}{Trek~S Palmer}, \bibinfo{person}{Darko Stefanovic}, {and}
  \bibinfo{person}{Dino~Dai Zovi}.} \bibinfo{year}{2003}\natexlab{}.
\newblock \showarticletitle{Randomized instruction set emulation to disrupt
  binary code injection attacks}. In \bibinfo{booktitle}{\emph{Proc. CCS}}.
  \bibinfo{pages}{281--289}.
\newblock


\bibitem[\protect\citeauthoryear{Belleville, Courouss\'{e}, Heydemann, and
  Charles}{Belleville et~al\mbox{.}}{2018}]%
        {10.1145/3281662}
\bibfield{author}{\bibinfo{person}{Nicolas Belleville}, \bibinfo{person}{Damien
  Courouss\'{e}}, \bibinfo{person}{Karine Heydemann}, {and}
  \bibinfo{person}{Henri-Pierre Charles}.} \bibinfo{year}{2018}\natexlab{}.
\newblock \showarticletitle{Automated Software Protection for the Masses
  Against Side-Channel Attacks}.
\newblock \bibinfo{journal}{\emph{ACM Trans. Archit. Code Optim.}}
  \bibinfo{volume}{15}, \bibinfo{number}{4}, Article \bibinfo{articleno}{47}
  (\bibinfo{date}{nov} \bibinfo{year}{2018}), \bibinfo{numpages}{27}~pages.
\newblock
\showISSN{1544-3566}
\urldef\tempurl%
\url{https://doi.org/10.1145/3281662}
\showDOI{\tempurl}


\bibitem[\protect\citeauthoryear{Bernstein}{Bernstein}{2005}]%
        {bernstein2005cache}
\bibfield{author}{\bibinfo{person}{Daniel~J Bernstein}.}
  \bibinfo{year}{2005}\natexlab{}.
\newblock \showarticletitle{Cache-timing attacks on AES}.
\newblock  (\bibinfo{year}{2005}).
\newblock


\bibitem[\protect\citeauthoryear{Bhatkar, DuVarney, and Sekar}{Bhatkar
  et~al\mbox{.}}{2003}]%
        {bhatkar03}
\bibfield{author}{\bibinfo{person}{Sandeep Bhatkar}, \bibinfo{person}{Daniel~C.
  DuVarney}, {and} \bibinfo{person}{R. Sekar}.}
  \bibinfo{year}{2003}\natexlab{}.
\newblock \showarticletitle{Address obfuscation: an efficient approach to
  combat a board range of memory error exploits}. In
  \bibinfo{booktitle}{\emph{Proceedings of the USENIX Security Symposium}}.
\newblock


\bibitem[\protect\citeauthoryear{Bhatkar, Sekar, and DuVarney}{Bhatkar
  et~al\mbox{.}}{2005}]%
        {bhatkar2005efficient}
\bibfield{author}{\bibinfo{person}{Sandeep Bhatkar}, \bibinfo{person}{Ron
  Sekar}, {and} \bibinfo{person}{Daniel~C DuVarney}.}
  \bibinfo{year}{2005}\natexlab{}.
\newblock \showarticletitle{Efficient techniques for comprehensive protection
  from memory error exploits}. In \bibinfo{booktitle}{\emph{Proceedings of the
  USENIX Security Symposium}}. \bibinfo{pages}{271--286}.
\newblock


\bibitem[\protect\citeauthoryear{Bian, Meng, and Zhang}{Bian
  et~al\mbox{.}}{2020}]%
        {bian2020minethrottle}
\bibfield{author}{\bibinfo{person}{Weikang Bian}, \bibinfo{person}{Wei Meng},
  {and} \bibinfo{person}{Mingxue Zhang}.} \bibinfo{year}{2020}\natexlab{}.
\newblock \showarticletitle{Minethrottle: Defending against wasm in-browser
  cryptojacking}. In \bibinfo{booktitle}{\emph{Proceedings of The Web
  Conference 2020}}. \bibinfo{pages}{3112--3118}.
\newblock


\bibitem[\protect\citeauthoryear{Brennan, Rosner, and Bultan}{Brennan
  et~al\mbox{.}}{2020}]%
        {brennan2020jit}
\bibfield{author}{\bibinfo{person}{Tegan Brennan}, \bibinfo{person}{Nicol{\'a}s
  Rosner}, {and} \bibinfo{person}{Tevfik Bultan}.}
  \bibinfo{year}{2020}\natexlab{}.
\newblock \showarticletitle{JIT Leaks: inducing timing side channels through
  just-in-time compilation}. In \bibinfo{booktitle}{\emph{2020 IEEE Symposium
  on Security and Privacy (SP)}}. IEEE, \bibinfo{pages}{1207--1222}.
\newblock


\bibitem[\protect\citeauthoryear{Bruschi, Cavallaro, and Lanzi}{Bruschi
  et~al\mbox{.}}{2007}]%
        {bruschi2007diversified}
\bibfield{author}{\bibinfo{person}{Danilo Bruschi}, \bibinfo{person}{Lorenzo
  Cavallaro}, {and} \bibinfo{person}{Andrea Lanzi}.}
  \bibinfo{year}{2007}\natexlab{}.
\newblock \showarticletitle{Diversified process replic{\ae} for defeating
  memory error exploits}. In \bibinfo{booktitle}{\emph{Proc. of the Int.
  Performance, Computing, and Communications Conference}}.
\newblock


\bibitem[\protect\citeauthoryear{Bryant}{Bryant}{2020}]%
        {bryant2020webassembly}
\bibfield{author}{\bibinfo{person}{David Bryant}.}
  \bibinfo{year}{2020}\natexlab{}.
\newblock \showarticletitle{Webassembly outside the browser: A new foundation
  for pervasive computing}. In \bibinfo{booktitle}{\emph{Proc. of ICWE 2020}}.
  \bibinfo{pages}{9--12}.
\newblock


\bibitem[\protect\citeauthoryear{Cabrera-Arteaga, Malivitsis, Vera-P{\'e}rez,
  Baudry, and Monperrus}{Cabrera-Arteaga et~al\mbox{.}}{2021}]%
        {CabreraArteaga2020CROWCD}
\bibfield{author}{\bibinfo{person}{Javier Cabrera-Arteaga},
  \bibinfo{person}{Orestis~Floros Malivitsis}, \bibinfo{person}{Oscar
  Vera-P{\'e}rez}, \bibinfo{person}{Benoit Baudry}, {and}
  \bibinfo{person}{Martin Monperrus}.} \bibinfo{year}{2021}\natexlab{}.
\newblock \showarticletitle{CROW: Code Diversification for WebAssembly}. In
  \bibinfo{booktitle}{\emph{MADWeb, NDSS 2021}}.
\newblock


\bibitem[\protect\citeauthoryear{Chew and Song}{Chew and Song}{2002}]%
        {Chew02mitigatingbuffer}
\bibfield{author}{\bibinfo{person}{Monica Chew} {and} \bibinfo{person}{Dawn
  Song}.} \bibinfo{year}{2002}\natexlab{}.
\newblock \bibinfo{booktitle}{\emph{Mitigating buffer overflows by operating
  system randomization}}.
\newblock \bibinfo{type}{{T}echnical {R}eport} CS-02-197.
  \bibinfo{institution}{Carnegie Mellon University}.
\newblock


\bibitem[\protect\citeauthoryear{Choy, Wong, Simon, and Rosenberg}{Choy
  et~al\mbox{.}}{2014}]%
        {choy2014hybrid}
\bibfield{author}{\bibinfo{person}{Sharon Choy}, \bibinfo{person}{Bernard
  Wong}, \bibinfo{person}{Gwendal Simon}, {and} \bibinfo{person}{Catherine
  Rosenberg}.} \bibinfo{year}{2014}\natexlab{}.
\newblock \showarticletitle{A hybrid edge-cloud architecture for reducing
  on-demand gaming latency}.
\newblock \bibinfo{journal}{\emph{Multimedia systems}} \bibinfo{volume}{20},
  \bibinfo{number}{5} (\bibinfo{year}{2014}), \bibinfo{pages}{503--519}.
\newblock


\bibitem[\protect\citeauthoryear{Cohen}{Cohen}{1993}]%
        {cohen93}
\bibfield{author}{\bibinfo{person}{Frederick~B Cohen}.}
  \bibinfo{year}{1993}\natexlab{}.
\newblock \showarticletitle{Operating system protection through program
  evolution.}
\newblock \bibinfo{journal}{\emph{Computers \& Security}} \bibinfo{volume}{12},
  \bibinfo{number}{6} (\bibinfo{year}{1993}), \bibinfo{pages}{565--584}.
\newblock


\bibitem[\protect\citeauthoryear{Coppens, De~Sutter, and Maebe}{Coppens
  et~al\mbox{.}}{2013}]%
        {coppens2013feedback}
\bibfield{author}{\bibinfo{person}{Bart Coppens}, \bibinfo{person}{Bjorn
  De~Sutter}, {and} \bibinfo{person}{Jonas Maebe}.}
  \bibinfo{year}{2013}\natexlab{}.
\newblock \showarticletitle{Feedback-driven binary code diversification}.
\newblock \bibinfo{journal}{\emph{ACM Transactions on Architecture and Code
  Optimization (TACO)}} \bibinfo{volume}{9}, \bibinfo{number}{4}
  (\bibinfo{year}{2013}), \bibinfo{pages}{1--26}.
\newblock


\bibitem[\protect\citeauthoryear{Cox, Evans, Filipi, Rowanhill, Hu, Davidson,
  Knight, Nguyen-Tuong, and Hiser}{Cox et~al\mbox{.}}{2006}]%
        {cox06}
\bibfield{author}{\bibinfo{person}{Benjamin Cox}, \bibinfo{person}{David
  Evans}, \bibinfo{person}{Adrian Filipi}, \bibinfo{person}{Jonathan
  Rowanhill}, \bibinfo{person}{Wei Hu}, \bibinfo{person}{Jack Davidson},
  \bibinfo{person}{John Knight}, \bibinfo{person}{Anh Nguyen-Tuong}, {and}
  \bibinfo{person}{Jason Hiser}.} \bibinfo{year}{2006}\natexlab{}.
\newblock \showarticletitle{N-variant systems: a secretless framework for
  security through diversity}. In \bibinfo{booktitle}{\emph{Proc. of USENIX
  Security Symposium}} (Vancouver, B.C., Canada)
  \emph{(\bibinfo{series}{USENIX-SS'06})}.
\newblock
\urldef\tempurl%
\url{http://dl.acm.org/citation.cfm?id=1267336.1267344}
\showURL{%
\tempurl}


\bibitem[\protect\citeauthoryear{Crane, Homescu, Brunthaler, Larsen, and
  Franz}{Crane et~al\mbox{.}}{2015}]%
        {crane2015thwarting}
\bibfield{author}{\bibinfo{person}{Stephen Crane}, \bibinfo{person}{Andrei
  Homescu}, \bibinfo{person}{Stefan Brunthaler}, \bibinfo{person}{Per Larsen},
  {and} \bibinfo{person}{Michael Franz}.} \bibinfo{year}{2015}\natexlab{}.
\newblock \showarticletitle{Thwarting Cache Side-Channel Attacks Through
  Dynamic Software Diversity}. In \bibinfo{booktitle}{\emph{NDSS}}.
  \bibinfo{pages}{8--11}.
\newblock


\bibitem[\protect\citeauthoryear{Davi, Liebchen, Sadeghi, Snow, and
  Monrose}{Davi et~al\mbox{.}}{2015}]%
        {davi2015isomeron}
\bibfield{author}{\bibinfo{person}{Lucas Davi}, \bibinfo{person}{Christopher
  Liebchen}, \bibinfo{person}{Ahmad-Reza Sadeghi}, \bibinfo{person}{Kevin~Z
  Snow}, {and} \bibinfo{person}{Fabian Monrose}.}
  \bibinfo{year}{2015}\natexlab{}.
\newblock \showarticletitle{Isomeron: Code Randomization Resilient to
  (Just-In-Time) Return-Oriented Programming}. In
  \bibinfo{booktitle}{\emph{NDSS}}.
\newblock


\bibitem[\protect\citeauthoryear{Forrest, Somayaji, and Ackley}{Forrest
  et~al\mbox{.}}{1997}]%
        {forrest97}
\bibfield{author}{\bibinfo{person}{Stephanie Forrest}, \bibinfo{person}{Anil
  Somayaji}, {and} \bibinfo{person}{David~H Ackley}.}
  \bibinfo{year}{1997}\natexlab{}.
\newblock \showarticletitle{Building diverse computer systems}. In
  \bibinfo{booktitle}{\emph{Proceedings. The Sixth Workshop on Hot Topics in
  Operating Systems}}. IEEE, \bibinfo{pages}{67--72}.
\newblock


\bibitem[\protect\citeauthoryear{Haas, Rossberg, Schuff, Titzer, Holman,
  Gohman, Wagner, Zakai, and Bastien}{Haas et~al\mbox{.}}{2017}]%
        {haas2017bringing}
\bibfield{author}{\bibinfo{person}{Andreas Haas}, \bibinfo{person}{Andreas
  Rossberg}, \bibinfo{person}{Derek~L Schuff}, \bibinfo{person}{Ben~L Titzer},
  \bibinfo{person}{Michael Holman}, \bibinfo{person}{Dan Gohman},
  \bibinfo{person}{Luke Wagner}, \bibinfo{person}{Alon Zakai}, {and}
  \bibinfo{person}{JF Bastien}.} \bibinfo{year}{2017}\natexlab{}.
\newblock \showarticletitle{Bringing the web up to speed with {WebAssembly}}.
  In \bibinfo{booktitle}{\emph{Proceedings of the 38th ACM SIGPLAN Conference
  on Programming Language Design and Implementation}}.
  \bibinfo{pages}{185--200}.
\newblock


\bibitem[\protect\citeauthoryear{Hickey}{Hickey}{2018}]%
        {FastlyWasm}
\bibfield{author}{\bibinfo{person}{Pat Hickey}.}
  \bibinfo{year}{2018}\natexlab{}.
\newblock \bibinfo{booktitle}{\emph{Announcing Lucet: Fastly’s native
  WebAssembly compiler and runtime}}.
\newblock \bibinfo{type}{{T}echnical {R}eport}.
\newblock
\urldef\tempurl%
\url{https://www.fastly.com/blog/announcing-lucet-fastly-native-webassembly-compiler-runtime}
\showURL{%
\tempurl}


\bibitem[\protect\citeauthoryear{Hilbig, Lehmann, and Pradel}{Hilbig
  et~al\mbox{.}}{2021}]%
        {hilbig2021empirical}
\bibfield{author}{\bibinfo{person}{Aaron Hilbig}, \bibinfo{person}{Daniel
  Lehmann}, {and} \bibinfo{person}{Michael Pradel}.}
  \bibinfo{year}{2021}\natexlab{}.
\newblock \showarticletitle{An Empirical Study of Real-World WebAssembly
  Binaries: Security, Languages, Use Cases}. In
  \bibinfo{booktitle}{\emph{Proceedings of the Web Conference 2021}}.
  \bibinfo{pages}{2696--2708}.
\newblock


\bibitem[\protect\citeauthoryear{Jackson}{Jackson}{2012}]%
        {jackson}
\bibfield{author}{\bibinfo{person}{Todd Jackson}.}
  \bibinfo{year}{2012}\natexlab{}.
\newblock \emph{\bibinfo{title}{On the Design, Implications, and Effects of
  Implementing Software Diversity for Security}}.
\newblock \bibinfo{thesistype}{Ph.D. Dissertation}. \bibinfo{school}{University
  of California, Irvine}.
\newblock


\bibitem[\protect\citeauthoryear{Jackson, Wimmer, and Franz}{Jackson
  et~al\mbox{.}}{2010}]%
        {jackson2010multi}
\bibfield{author}{\bibinfo{person}{Todd Jackson}, \bibinfo{person}{Christian
  Wimmer}, {and} \bibinfo{person}{Michael Franz}.}
  \bibinfo{year}{2010}\natexlab{}.
\newblock \showarticletitle{Multi-variant program execution for vulnerability
  detection and analysis}. In \bibinfo{booktitle}{\emph{Proceedings of the
  Sixth Annual Workshop on Cyber Security and Information Intelligence
  Research}}. \bibinfo{pages}{1--4}.
\newblock


\bibitem[\protect\citeauthoryear{Jacobsson and W{\aa}hsl{\'e}n}{Jacobsson and
  W{\aa}hsl{\'e}n}{2018}]%
        {1244493Jacobsson}
\bibfield{author}{\bibinfo{person}{Martin Jacobsson} {and}
  \bibinfo{person}{Jonas W{\aa}hsl{\'e}n}.} \bibinfo{year}{2018}\natexlab{}.
\newblock \showarticletitle{Virtual machine execution for wearables based on
  webassembly}. In \bibinfo{booktitle}{\emph{EAI International Conference on
  Body Area Networks}}. Springer, Cham, \bibinfo{pages}{381--389}.
\newblock


\bibitem[\protect\citeauthoryear{Johnson, Thien, Alhessi, Narayan, Brown,
  Lerner, McMullen, Savage, and Stefan}{Johnson et~al\mbox{.}}{2021}]%
        {johnson2021}
\bibfield{author}{\bibinfo{person}{Evan Johnson}, \bibinfo{person}{David
  Thien}, \bibinfo{person}{Yousef Alhessi}, \bibinfo{person}{Shravan Narayan},
  \bibinfo{person}{Fraser Brown}, \bibinfo{person}{Sorin Lerner},
  \bibinfo{person}{Tyler McMullen}, \bibinfo{person}{Stefan Savage}, {and}
  \bibinfo{person}{Deian Stefan}.} \bibinfo{year}{2021}\natexlab{}.
\newblock \showarticletitle{SFI safety for native-compiled Wasm}.
\newblock \bibinfo{journal}{\emph{NDSS. Internet Society}}
  (\bibinfo{year}{2021}).
\newblock


\bibitem[\protect\citeauthoryear{Kc, Keromytis, and Prevelakis}{Kc
  et~al\mbox{.}}{2003}]%
        {Kc03}
\bibfield{author}{\bibinfo{person}{Gaurav~S. Kc}, \bibinfo{person}{Angelos~D.
  Keromytis}, {and} \bibinfo{person}{Vassilis Prevelakis}.}
  \bibinfo{year}{2003}\natexlab{}.
\newblock \showarticletitle{Countering code-injection attacks with
  instruction-set randomization}. In \bibinfo{booktitle}{\emph{Proc. of CCS}}.
  \bibinfo{pages}{272--280}.
\newblock


\bibitem[\protect\citeauthoryear{Kim, Kwon, Sumner, Zhang, and Xu}{Kim
  et~al\mbox{.}}{2015}]%
        {Kim2015}
\bibfield{author}{\bibinfo{person}{Dohyeong Kim}, \bibinfo{person}{Yonghwi
  Kwon}, \bibinfo{person}{William~N. Sumner}, \bibinfo{person}{Xiangyu Zhang},
  {and} \bibinfo{person}{Dongyan Xu}.} \bibinfo{year}{2015}\natexlab{}.
\newblock \showarticletitle{Dual Execution for On the Fly Fine Grained
  Execution Comparison}.
\newblock \bibinfo{journal}{\emph{SIGPLAN Not.}} (\bibinfo{year}{2015}).
\newblock


\bibitem[\protect\citeauthoryear{Koning, Bos, and Giuffrida}{Koning
  et~al\mbox{.}}{2016}]%
        {koning2016secure}
\bibfield{author}{\bibinfo{person}{Koen Koning}, \bibinfo{person}{Herbert Bos},
  {and} \bibinfo{person}{Cristiano Giuffrida}.}
  \bibinfo{year}{2016}\natexlab{}.
\newblock \showarticletitle{Secure and efficient multi-variant execution using
  hardware-assisted process virtualization}. In \bibinfo{booktitle}{\emph{2016
  46th Annual IEEE/IFIP International Conference on Dependable Systems and
  Networks (DSN)}}. IEEE, \bibinfo{pages}{431--442}.
\newblock


\bibitem[\protect\citeauthoryear{Lee, Kang, Jang, and Kang}{Lee
  et~al\mbox{.}}{2021}]%
        {lee2021savior}
\bibfield{author}{\bibinfo{person}{Seongman Lee}, \bibinfo{person}{Hyeonwoo
  Kang}, \bibinfo{person}{Jinsoo Jang}, {and} \bibinfo{person}{Brent~Byunghoon
  Kang}.} \bibinfo{year}{2021}\natexlab{}.
\newblock \showarticletitle{SaVioR: Thwarting Stack-Based Memory Safety
  Violations by Randomizing Stack Layout}.
\newblock \bibinfo{journal}{\emph{IEEE Transactions on Dependable and Secure
  Computing}} (\bibinfo{year}{2021}).
\newblock
\urldef\tempurl%
\url{https://ieeexplore.ieee.org/iel7/8858/4358699/09369900.pdf}
\showURL{%
\tempurl}


\bibitem[\protect\citeauthoryear{Lehmann, Kinder, and Pradel}{Lehmann
  et~al\mbox{.}}{2020}]%
        {usenixWasm2020}
\bibfield{author}{\bibinfo{person}{Daniel Lehmann}, \bibinfo{person}{Johannes
  Kinder}, {and} \bibinfo{person}{Michael Pradel}.}
  \bibinfo{year}{2020}\natexlab{}.
\newblock \showarticletitle{Everything Old is New Again: Binary Security of
  WebAssembly}. In \bibinfo{booktitle}{\emph{29th USENIX Security Symposium
  (USENIX Security 20)}}. \bibinfo{publisher}{USENIX Association}.
\newblock


\bibitem[\protect\citeauthoryear{Lettner, Song, Park, Larsen, Volckaert, and
  Franz}{Lettner et~al\mbox{.}}{2018}]%
        {lettner2018partisan}
\bibfield{author}{\bibinfo{person}{Julian Lettner}, \bibinfo{person}{Dokyung
  Song}, \bibinfo{person}{Taemin Park}, \bibinfo{person}{Per Larsen},
  \bibinfo{person}{Stijn Volckaert}, {and} \bibinfo{person}{Michael Franz}.}
  \bibinfo{year}{2018}\natexlab{}.
\newblock \showarticletitle{PartiSan: fast and flexible sanitization via
  run-time partitioning}. In \bibinfo{booktitle}{\emph{International Symposium
  on Research in Attacks, Intrusions, and Defenses}}. Springer,
  \bibinfo{pages}{403--422}.
\newblock


\bibitem[\protect\citeauthoryear{Liljestrand, Nyman, Gunn, Ekberg, and
  Asokan}{Liljestrand et~al\mbox{.}}{2021}]%
        {liljestrand2021pacstack}
\bibfield{author}{\bibinfo{person}{Hans Liljestrand}, \bibinfo{person}{Thomas
  Nyman}, \bibinfo{person}{Lachlan~J Gunn}, \bibinfo{person}{Jan-Erik Ekberg},
  {and} \bibinfo{person}{N Asokan}.} \bibinfo{year}{2021}\natexlab{}.
\newblock \showarticletitle{PACStack: an Authenticated Call Stack}. In
  \bibinfo{booktitle}{\emph{30th USENIX Security Symposium (USENIX Security
  21)}}.
\newblock


\bibitem[\protect\citeauthoryear{Lu, Xu, Song, Kim, and Lee}{Lu
  et~al\mbox{.}}{2018}]%
        {lu2018stopping}
\bibfield{author}{\bibinfo{person}{Kangjie Lu}, \bibinfo{person}{Meng Xu},
  \bibinfo{person}{Chengyu Song}, \bibinfo{person}{Taesoo Kim}, {and}
  \bibinfo{person}{Wenke Lee}.} \bibinfo{year}{2018}\natexlab{}.
\newblock \showarticletitle{Stopping memory disclosures via diversification and
  replicated execution}.
\newblock \bibinfo{journal}{\emph{IEEE Transactions on Dependable and Secure
  Computing}} (\bibinfo{year}{2018}).
\newblock


\bibitem[\protect\citeauthoryear{Mann and Whitney}{Mann and Whitney}{1947}]%
        {mann1947}
\bibfield{author}{\bibinfo{person}{H.~B. Mann} {and} \bibinfo{person}{D.~R.
  Whitney}.} \bibinfo{year}{1947}\natexlab{}.
\newblock \showarticletitle{On a Test of Whether one of Two Random Variables is
  Stochastically Larger than the Other}.
\newblock \bibinfo{journal}{\emph{Ann. Math. Statist.}} \bibinfo{volume}{18},
  \bibinfo{number}{1} (\bibinfo{date}{03} \bibinfo{year}{1947}),
  \bibinfo{pages}{50--60}.
\newblock
\urldef\tempurl%
\url{https://doi.org/10.1214/aoms/1177730491}
\showDOI{\tempurl}


\bibitem[\protect\citeauthoryear{Maurer and Brumley}{Maurer and
  Brumley}{2012}]%
        {maurer2012tachyon}
\bibfield{author}{\bibinfo{person}{Matthew Maurer} {and} \bibinfo{person}{David
  Brumley}.} \bibinfo{year}{2012}\natexlab{}.
\newblock \showarticletitle{TACHYON: Tandem execution for efficient live patch
  testing}. In \bibinfo{booktitle}{\emph{21st USENIX Security Symposium (USENIX
  Security 12)}}. \bibinfo{pages}{617--630}.
\newblock


\bibitem[\protect\citeauthoryear{{Mendki}}{{Mendki}}{2020}]%
        {pMendkiServerless}
\bibfield{author}{\bibinfo{person}{P. {Mendki}}.}
  \bibinfo{year}{2020}\natexlab{}.
\newblock \showarticletitle{Evaluating Webassembly Enabled Serverless Approach
  for Edge Computing}. In \bibinfo{booktitle}{\emph{2020 IEEE Cloud Summit}}.
  \bibinfo{pages}{161--166}.
\newblock
\urldef\tempurl%
\url{https://doi.org/10.1109/IEEECloudSummit48914.2020.00031}
\showDOI{\tempurl}


\bibitem[\protect\citeauthoryear{Narayan, Disselkoen, Moghimi, Cauligi,
  Johnson, Gang, Vahldiek-Oberwagner, Sahita, Shacham, Tullsen,
  et~al\mbox{.}}{Narayan et~al\mbox{.}}{2021}]%
        {Narayan2021Swivel}
\bibfield{author}{\bibinfo{person}{Shravan Narayan}, \bibinfo{person}{Craig
  Disselkoen}, \bibinfo{person}{Daniel Moghimi}, \bibinfo{person}{Sunjay
  Cauligi}, \bibinfo{person}{Evan Johnson}, \bibinfo{person}{Zhao Gang},
  \bibinfo{person}{Anjo Vahldiek-Oberwagner}, \bibinfo{person}{Ravi Sahita},
  \bibinfo{person}{Hovav Shacham}, \bibinfo{person}{Dean Tullsen},
  {et~al\mbox{.}}} \bibinfo{year}{2021}\natexlab{}.
\newblock \showarticletitle{Swivel: Hardening WebAssembly against Spectre}. In
  \bibinfo{booktitle}{\emph{USENIX Security Symposium}}.
\newblock


\bibitem[\protect\citeauthoryear{O'Donnell and Sethu}{O'Donnell and
  Sethu}{2004}]%
        {o2004achieving}
\bibfield{author}{\bibinfo{person}{Adam~J O'Donnell} {and}
  \bibinfo{person}{Harish Sethu}.} \bibinfo{year}{2004}\natexlab{}.
\newblock \showarticletitle{On achieving software diversity for improved
  network security using distributed coloring algorithms}. In
  \bibinfo{booktitle}{\emph{Proceedings of the 11th ACM conference on Computer
  and communications security}}. \bibinfo{pages}{121--131}.
\newblock


\bibitem[\protect\citeauthoryear{{\"O}sterlund, Koning, Olivier, Barbalace,
  Bos, and Giuffrida}{{\"O}sterlund et~al\mbox{.}}{2019}]%
        {osterlund2019kmvx}
\bibfield{author}{\bibinfo{person}{Sebastian {\"O}sterlund},
  \bibinfo{person}{Koen Koning}, \bibinfo{person}{Pierre Olivier},
  \bibinfo{person}{Antonio Barbalace}, \bibinfo{person}{Herbert Bos}, {and}
  \bibinfo{person}{Cristiano Giuffrida}.} \bibinfo{year}{2019}\natexlab{}.
\newblock \showarticletitle{kMVX: Detecting kernel information leaks with
  multi-variant execution}. In \bibinfo{booktitle}{\emph{ASPLOS}}.
\newblock


\bibitem[\protect\citeauthoryear{Rane, Lin, and Tiwari}{Rane
  et~al\mbox{.}}{2015}]%
        {rane2015raccoon}
\bibfield{author}{\bibinfo{person}{Ashay Rane}, \bibinfo{person}{Calvin Lin},
  {and} \bibinfo{person}{Mohit Tiwari}.} \bibinfo{year}{2015}\natexlab{}.
\newblock \showarticletitle{Raccoon: Closing digital side-channels through
  obfuscated execution}. In \bibinfo{booktitle}{\emph{24th USENIX Security
  Symposium (USENIX Security 15)}}. \bibinfo{pages}{431--446}.
\newblock


\bibitem[\protect\citeauthoryear{Ryder}{Ryder}{1979}]%
        {ryder1979}
\bibfield{author}{\bibinfo{person}{Barbara~G Ryder}.}
  \bibinfo{year}{1979}\natexlab{}.
\newblock \showarticletitle{Constructing the call graph of a program}.
\newblock \bibinfo{journal}{\emph{IEEE Transactions on Software Engineering}}
  \bibinfo{number}{3} (\bibinfo{year}{1979}), \bibinfo{pages}{216--226}.
\newblock


\bibitem[\protect\citeauthoryear{Salamat, Gal, Jackson, Manivannan, Wagner, and
  Franz}{Salamat et~al\mbox{.}}{2007}]%
        {salamat2007stopping}
\bibfield{author}{\bibinfo{person}{Babak Salamat}, \bibinfo{person}{Andreas
  Gal}, \bibinfo{person}{Todd Jackson}, \bibinfo{person}{Karthik Manivannan},
  \bibinfo{person}{Gregor Wagner}, {and} \bibinfo{person}{Michael Franz}.}
  \bibinfo{year}{2007}\natexlab{}.
\newblock \bibinfo{booktitle}{\emph{Stopping Buffer Overflow Attacks at
  Run-Time: Simultaneous Multi-Variant Program Execution on a Multicore
  Processor}}.
\newblock \bibinfo{type}{{T}echnical {R}eport}. \bibinfo{institution}{Technical
  Report 07-13, School of Information and Computer Sciences, UCIrvine}.
\newblock


\bibitem[\protect\citeauthoryear{Salamat, Jackson, Gal, and Franz}{Salamat
  et~al\mbox{.}}{2009}]%
        {salamat2009orchestra}
\bibfield{author}{\bibinfo{person}{Babak Salamat}, \bibinfo{person}{Todd
  Jackson}, \bibinfo{person}{Andreas Gal}, {and} \bibinfo{person}{Michael
  Franz}.} \bibinfo{year}{2009}\natexlab{}.
\newblock \showarticletitle{Orchestra: intrusion detection using parallel
  execution and monitoring of program variants in user-space}. In
  \bibinfo{booktitle}{\emph{Proceedings of the 4th ACM European conference on
  Computer systems}}. \bibinfo{pages}{33--46}.
\newblock


\bibitem[\protect\citeauthoryear{Salamat, Jackson, Wagner, Wimmer, and
  Franz}{Salamat et~al\mbox{.}}{2011}]%
        {SalamatJWWF11}
\bibfield{author}{\bibinfo{person}{Babak Salamat}, \bibinfo{person}{Todd
  Jackson}, \bibinfo{person}{Gregor Wagner}, \bibinfo{person}{Christian
  Wimmer}, {and} \bibinfo{person}{Michael Franz}.}
  \bibinfo{year}{2011}\natexlab{}.
\newblock \showarticletitle{Runtime Defense against Code Injection Attacks
  Using Replicated Execution}.
\newblock \bibinfo{journal}{\emph{{IEEE} Trans. Dependable Secur. Comput.}}
  \bibinfo{volume}{8}, \bibinfo{number}{4} (\bibinfo{year}{2011}),
  \bibinfo{pages}{588--601}.
\newblock
\urldef\tempurl%
\url{https://doi.org/10.1109/TDSC.2011.18}
\showDOI{\tempurl}


\bibitem[\protect\citeauthoryear{Shillaker and Pietzuch}{Shillaker and
  Pietzuch}{2020}]%
        {shillaker2020faasm}
\bibfield{author}{\bibinfo{person}{Simon Shillaker} {and}
  \bibinfo{person}{Peter Pietzuch}.} \bibinfo{year}{2020}\natexlab{}.
\newblock \showarticletitle{Faasm: Lightweight isolation for efficient stateful
  serverless computing}. In \bibinfo{booktitle}{\emph{USENIX Annual Technical
  Conference}}. \bibinfo{pages}{419--433}.
\newblock


\bibitem[\protect\citeauthoryear{Silvanovich}{Silvanovich}{2018}]%
        {ChromeZero}
\bibfield{author}{\bibinfo{person}{Natalie Silvanovich}.}
  \bibinfo{year}{2018}\natexlab{}.
\newblock \bibinfo{booktitle}{\emph{The Problems and Promise of WebAssembly}}.
\newblock \bibinfo{type}{{T}echnical {R}eport}.
\newblock
\urldef\tempurl%
\url{https://googleprojectzero.blogspot.com/2018/08/the-problems-and-promise-of-webassembly.html}
\showURL{%
\tempurl}


\bibitem[\protect\citeauthoryear{Sti{\'e}venart and De~Roover}{Sti{\'e}venart
  and De~Roover}{2020}]%
        {stievenart2020compositional}
\bibfield{author}{\bibinfo{person}{Quentin Sti{\'e}venart} {and}
  \bibinfo{person}{Coen De~Roover}.} \bibinfo{year}{2020}\natexlab{}.
\newblock \showarticletitle{Compositional Information Flow Analysis for
  WebAssembly Programs}. In \bibinfo{booktitle}{\emph{2020 IEEE 20th
  International Working Conference on Source Code Analysis and Manipulation
  (SCAM)}}. IEEE, \bibinfo{pages}{13--24}.
\newblock


\bibitem[\protect\citeauthoryear{Taleb, Samdanis, Mada, Flinck, Dutta, and
  Sabella}{Taleb et~al\mbox{.}}{2017}]%
        {taleb2017multi}
\bibfield{author}{\bibinfo{person}{Tarik Taleb}, \bibinfo{person}{Konstantinos
  Samdanis}, \bibinfo{person}{Badr Mada}, \bibinfo{person}{Hannu Flinck},
  \bibinfo{person}{Sunny Dutta}, {and} \bibinfo{person}{Dario Sabella}.}
  \bibinfo{year}{2017}\natexlab{}.
\newblock \showarticletitle{On Multi-Access Edge Computing: A Survey of the
  Emerging 5G Network Edge Cloud Architecture and Orchestration}.
\newblock \bibinfo{journal}{\emph{IEEE Comm. Surveys \& Tutorials}}
  \bibinfo{volume}{19}, \bibinfo{number}{3} (\bibinfo{year}{2017}).
\newblock


\bibitem[\protect\citeauthoryear{Varda}{Varda}{2018}]%
        {CloudflareWasm}
\bibfield{author}{\bibinfo{person}{Kenton Varda}.}
  \bibinfo{year}{2018}\natexlab{}.
\newblock \bibinfo{booktitle}{\emph{WebAssembly on Cloudflare Workers}}.
\newblock \bibinfo{type}{{T}echnical {R}eport}.
\newblock
\urldef\tempurl%
\url{https://blog.cloudflare.com/webassembly-on-cloudflare-workers/}
\showURL{%
\tempurl}


\bibitem[\protect\citeauthoryear{Volckaert, Coppens, and De~Sutter}{Volckaert
  et~al\mbox{.}}{2015}]%
        {volckaert2015cloning}
\bibfield{author}{\bibinfo{person}{Stijn Volckaert}, \bibinfo{person}{Bart
  Coppens}, {and} \bibinfo{person}{Bjorn De~Sutter}.}
  \bibinfo{year}{2015}\natexlab{}.
\newblock \showarticletitle{Cloning your gadgets: Complete ROP attack immunity
  with multi-variant execution}.
\newblock \bibinfo{journal}{\emph{IEEE Transactions on Dependable and Secure
  Computing}} \bibinfo{volume}{13}, \bibinfo{number}{4} (\bibinfo{year}{2015}).
\newblock


\bibitem[\protect\citeauthoryear{Voulimeneas, Song, Larsen, Franz, and
  Volckaert}{Voulimeneas et~al\mbox{.}}{2021}]%
        {voulimeneas2021dmvx}
\bibfield{author}{\bibinfo{person}{Alexios Voulimeneas},
  \bibinfo{person}{Dokyung Song}, \bibinfo{person}{Per Larsen},
  \bibinfo{person}{Michael Franz}, {and} \bibinfo{person}{Stijn Volckaert}.}
  \bibinfo{year}{2021}\natexlab{}.
\newblock \showarticletitle{dMVX: Secure and Efficient Multi-Variant Execution
  in a Distributed Setting}. In \bibinfo{booktitle}{\emph{Proceedings of the
  14th European Workshop on Systems Security}}. \bibinfo{pages}{41--47}.
\newblock


\bibitem[\protect\citeauthoryear{Xu, Solihin, and Shen}{Xu
  et~al\mbox{.}}{2020}]%
        {xu2020merr}
\bibfield{author}{\bibinfo{person}{Yuanchao Xu}, \bibinfo{person}{Yan Solihin},
  {and} \bibinfo{person}{Xipeng Shen}.} \bibinfo{year}{2020}\natexlab{}.
\newblock \showarticletitle{Merr: Improving security of persistent memory
  objects via efficient memory exposure reduction and randomization}. In
  \bibinfo{booktitle}{\emph{Proc. of ASPLOS}}. \bibinfo{pages}{987--1000}.
\newblock
\urldef\tempurl%
\url{https://dl.acm.org/doi/pdf/10.1145/3373376.3378492}
\showURL{%
\tempurl}


\end{thebibliography}
\bibliographystyle{ACM-Reference-Format}

\appendix
\section{Dispatcher function code}
\label{dispatcher_code}

\lstset{
    language=llvm,
    style=nccode,
    basicstyle=\footnotesize\ttfamily,
    columns=fullflexible,
    breaklines=true,
    numbers=none,
    stepnumber=1,
    float
}
\begin{code}
\scriptsize
\noindent\begin{minipage}[b]{\linewidth}
    \begin{minipage}[t]{1\linewidth}
    \begin{lstlisting}[escapeinside={(*}{*)}]
define internal i32 @b64_byte2urlsafe_char(i32 %0) {
    entry:
      %1 = call i32 @discriminate(i32 3)
      switch i32 %1, label %end [ i32 0, label %case_43_ i32 1, label %case_44_
      ]
    case_43_:                ; preds = %entry
      %2 = call i32 @b64_byte_to_urlsafe_char_43_(%0)
      ret i32 %2
    case_44_:                ; preds = %entry
      %3 = <body of b64_byte_to_urlsafe_char_44_>
      ret i32 %3
    end:                                              ; preds = %entry
      %4 = call i32 @b64_byte2urlsafe_char_original(%0)
      ret i32 %4
}
        \end{lstlisting}
    \end{minipage}
    \noindent\rule{\linewidth}{0.4pt}
    \captionof{lstlisting}{Dispatcher function embedded in the multivariant binary of the bin2base64 endpoint of libsodium, which corresponds to the rightmost green node in \autoref{multivariant}.}\label{listing:multivariant_template}
\end{minipage}
\vspace{-7mm}
\end{code}

\section{Multivariant Binary Execution at the Edge}
\label{app:instance}

% What really is executed is the x86 code
When a WebAssembly binary is deployed on an edge platform, it is translated to machine code on the fly.
For our experiment, we deploy on the production edge nodes of Fastly. This edge computing platform uses Lucet, a native WebAssembly compiler and runtime, to compile and run the deployed Wasm binary \footnote{\url{https://github.com/bytecodealliance/lucet}}.
Lucet generates x86 machine code and ensures that the generated machine code executes inside a secure sandbox, controlling memory isolation.

\begin{figure}
\centering
  \includegraphics[trim=22 15 5 10,clip,width=0.8\linewidth]{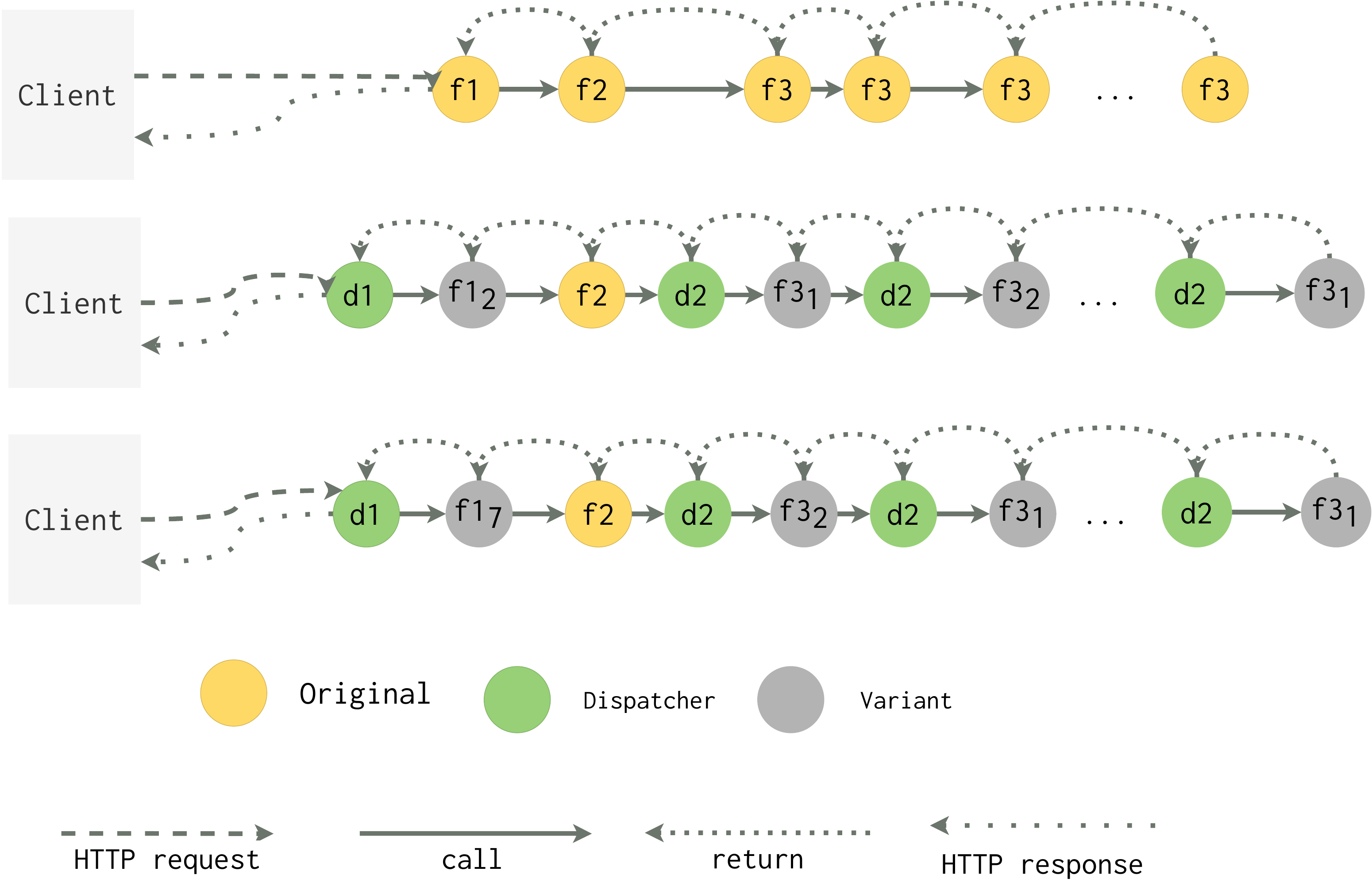}
  \caption{Top: an execution trace for the  \texttt{bin2base64} endpoint. Middle and bottom: two different execution traces for the multivariant \texttt{bin2base64}, exhibited by two different requests with exactly the same input.}
  \label{http:workflow}
\vspace{-5mm}
\end{figure}

% How it works
\autoref{http:workflow} illustrates  the runtime behavior of the original and the multivariant binary,  when deployed on an Edge node.
The top most diagram illustrates the execution trace for the  original of the endpoint \texttt{bin2base64}.
When the HTTP request with the input \texttt{"HelloWorld!"} is received, it invokes functions $f1$, $f2$ followed by 27 recursive calls of function $f3$. Then, the endpoint sends the result \\ \texttt{"0x000xccv0x10x00b3Jsx130x000x00 0x00xpopAHRvdGE="} of its base64 encoding in an HTTP response.

The two diagrams at the bottom of \autoref{http:workflow} illustrate two executions traces observed through two different requests to the endpoint \texttt{bin2base64}.
In the first case, the request first triggers the invocation of dispatcher $d1$, which randomly decides to invoke the variant $f1_2$; then $f2$, which has not been diversified by \tool, is invoked; then the recursive invocations to $f3$ are replaced by iterations over the execution of dispatcher $d2$ followed by a random choice of variants of $f3$. Eventually the result is computed and sent back as an HTTP response. 
The second execution trace of the multivariant binary shows the same sequence of dispatcher and function calls as the previous trace, and also shows that for a different requests, the variants of $f1$ and $f3$ are different.

The key insights from these figures are as follows. First, from a client's point of view, a request to the original or to a multivariant endpoint, is completely transparent. Clients send the same data, receive the same result, through the same protocol, in both cases.
Second, this figure shows that, at runtime, the execution paths for the same endpoint are different from one execution to another, and that this randomization process results from multiple random choices among function variants, made through the execution of the endpoint.
\section{Variants preservation}

During our experiments, we checked for code diversity preservation after compilation. In this work, diversity is introduced through transformation on  \wasm code, which is then compiled by the Lucet compiler. Compilation might perform some normalization and optimization passes when translating from \wasm to machine code. 
Thus, some variants synthesized by \tool might not be preserved, i.e., Lucet could generate the same machine code for two \wasm variants. To assess this potential effect, we compare the level of code diversity among the \wasm variants and among the machine code variants produced by Lucet. This experiment reveals that the translation to machine code preserves a high ratio of function variants, i.e., approx 96\% of the generated variants are preserved. This result also indicates that the machine code variants preserve the potential for large numbers of possible execution paths.

\end{document}